\documentclass[onefignum,onetabnum]{siamart190516}

\usepackage{amsfonts}
\usepackage{graphicx}
\usepackage{epstopdf}
\usepackage{algorithmic}
\ifpdf
   \DeclareGraphicsExtensions{.eps,.pdf,.png,.jpg}
  \else
   \DeclareGraphicsExtensions{.eps}
\fi

\newsiamremark{remark}{Remark}
\newsiamremark{hypothesis}{Hypothesis}
\crefname{hypothesis}{Hypothesis}{Hypotheses}
\newsiamthm{claim}{Claim}
\newsiamthm{property}{Property}
\usepackage{amssymb}
\usepackage{amsmath}
\usepackage{xcolor}
\usepackage{mathrsfs}
\usepackage{enumitem}
\usepackage{scalerel} %To scale text in \geqt etc.
\usepackage{bbm} %To use indicator function
\usepackage[normalem]{ulem} %to strikeout text (\sout)
\usepackage{mathtools} %for dcases
\usepackage{float} %to force location of figure[H]
\usepackage{enumitem} %to suppress spaces between items in a list
\usepackage{comment} %to comment out parts of the text
\usepackage{amsopn}
\usepackage{blkarray} %to put labels on column rows/cols

\newcommand{\Aut}{\operatorname{Aut}}
\newcommand{\supp}{\operatorname{supp}}
\newcommand{\geqt}{\geq_{\scalerel*{\mathcal{T}}{{*}}}}
\newcommand{\leqt}{\leq_{\scalerel*{\mathcal{T}}{{*}}}}
\newcommand{\lesst}{<_{\scalerel*{\mathcal{T}}{{*}}}}
\newcommand{\greatt}{>_{\scalerel*{\mathcal{T}}{{*}}}}
\newcommand{\geqp}{\geq_{\scalerel*{\mathcal{P}}{{*}}}}
\newcommand{\geqtsub}{\geq_{{\scaleto{\mathcal{T}}{3pt}}}}
\newcommand{\greattsub}{>_{{\scaleto{\mathcal{T}}{3pt}}}}

\headers{CONSTRUCTING LAPLACIAN MATRICES WITH SOULES VECTORS}{K. Devriendt, R. Lambiotte, and P. Van Mieghem}

\title{CONSTRUCTING LAPLACIAN MATRICES WITH SOULES VECTORS: INVERSE EIGENVALUE PROBLEM AND APPLICATIONS
}

\author{Karel Devriendt\thanks{Mathematical Institute, University of Oxford, Oxford UK and The Alan Turing Institute, London UK.}\and Renaud Lambiotte\thanks{Mathematical Institute, University of Oxford, Oxford UK.}\and Piet Van Mieghem\thanks{Department of Electrical Engineering, Mathematics and Computer Science, Delft University of Technology, Delft, Netherlands.}}

\begin{document}

\maketitle

% REQUIRED
\begin{abstract}
The symmetric nonnegative inverse eigenvalue problem (SNIEP) asks which sets of numbers (counting multiplicities) can be the eigenvalues of a symmetric matrix with nonnegative entries. While examples of such matrices are abundant in linear algebra and various applications, this question is still open for matrices of dimension $N\geq 5$. One of the approaches to solve the SNIEP was proposed by George W. Soules \cite{Soules}, relying on a specific type of eigenvectors (Soules vectors) to derive sufficient conditions for this problem. Elsner et al. \cite{Elsner} later showed a canonical way to construct all Soules vectors, based on binary rooted trees. While Soules vectors are typically treated as a totally ordered set of vectors, we propose in this article to consider a relaxed alternative: a partially ordered set of Soules vectors. We show that this perspective enables a more complete characterization of the sufficient conditions for the SNIEP. In particular, we show that the set of eigenvalues that satisfy these sufficient conditions is a convex cone, with symmetries corresponding to the automorphisms of the binary rooted tree from which the Soules vectors were constructed. As a second application, we show how Soules vectors can be used to construct graph Laplacian matrices with a given spectrum and describe a number of interesting connections with the concepts of hierarchical random graphs, equitable partitions and effective resistance.
\end{abstract}

% REQUIRED

% REQUIRED

%%%%%%
%BODY
%%%%%%
\section{Introduction}\label{S1: Intro} A central topic in linear algebra and matrix theory is the problem of characterizing the possible eigenvalues for different classes of matrices. These types of problems, which are often referred to as inverse eigenvalue problems\footnote{These problems are sometimes referred to as inverse spectrum problems instead, to stress the fact that all eigenvalues need to be characterized at the same time.}, trace back to the work of Kolmogorov, Sule\u{\i}manova and Perfect in the mid-twentieth century \cite{Egleston}. One important class of matrices are symmetric matrices with nonnegative entries, which feature prominently in linear algebra, for instance as {stochastic matrices} describing the transition probabilities of a Markov chain, {adjacency matrices} whose (typically sparse) entries encode the structure of a graph, or {distance matrices} that describe the pairwise distances of a finite metric space. For symmetric nonnegative matrices (SNN), the inverse eigenvalue problem is commonly known as the \emph{symmetric nonnegative inverse eigenvalue problem} (SNIEP), and is stated as follows:
\begin{align*}
&\text{\emph{Which sets of numbers can be the eigenvalues of a symmetric nonnegative matrix?}}
\end{align*}
Since its proposal by Miroslav Fiedler \cite{Fiedler} in 1974, several approaches have been developed to address the SNIEP, yielding different necessary and sufficient conditions for realizable sets \cite{Egleston},\cite{Marijuan}. However, the problem remains unsolved for dimension $N\geq 5$.
\\
In this article, we focus on an approach proposed by George W. Soules \cite{Soules} in 1983 to address the SNIEP in which he introduced a set of ordered vectors $r_1\geq \dots \geq r_N$ with the remarkable property that $\sum_{n=1}^N \lambda_nr_nr_n^T$ is an SNN matrix whenever $\lambda_1\geq\dots\geq\lambda_N\geq 0$ is satisfied. In subsequent work, Elsner et al. \cite{Elsner} studied all possible sets of vectors $\lbrace r_n\rbrace$ with the above property - naming them \emph{Soules vectors} - and gave a canonical way to construct these vectors from a binary rooted tree $\mathcal{T}$. The main contribution of this article is to reformulate Soules' approach in terms of \emph{partially ordered Soules vectors and eigenvalues}, replacing the total order used in \cite{Soules},\cite{Elsner} and subsequent work. We show that this reformulation is more natural for Soules vectors, given their construction from binary rooted trees, and that it enables a more complete characterization of the sufficient conditions for the SNIEP derived from a set of Soules vectors. In particular, we show that these sufficient conditions determine a convex cone which has symmetries related to the automorphism group of the corresponding tree $\mathcal{T}$. We additionally discuss the entries of matrices with Soules eigenvectors, revealing that these matrices posses a certain block structure. A second important contribution of this article is to show how Soules vectors can be used to construct a graph Laplacian matrix with any given set of eigenvalues. 
Furthermore, we discuss how the structure of the Soules vectors is reflected in the properties of the constructed graph, in particular the effective resistance and equitable partitions.
\\
\\
Section \ref{SS1.1: Related work} below discusses related work, both related to the SNIEP as to the theory of Soules vectors. In Section \ref{S2: Preliminaries}, we introduce the relevant mathematical concepts and notations. In Section \ref{S3: Soules bases}, we define totally ordered Soules bases and our relaxation to partially ordered Soules bases, as well as discussing some properties of matrices with Soules eigenvectors. Section \ref{A2.2: example} provides a detailed numerical example of the construction of Soules vectors and matrices with Soules eigenvectors. Section \ref{S4: SNIEP} discusses the symmetric nonnegative inverse eigenvalue problem and the methods to derive necessary and sufficient conditions. In Section \ref{S5: POSB and SNIEP}, sufficient conditions for the SNIEP are derived from Soules' approach and the properties of these conditions are discussed in Section \ref{SS5.1: Properties of set}. Section \ref{S7: Laplacian} introduces the construction of Laplacian matrices using Soules vectors and discusses some properties of these matrices and corresponding graphs. Finally, Section \ref{S8: Conclusion} concludes and summarizes the article.
%%%%%%%%%%%%
%
%
%%%%%%%%%%%%%
%
%
%%%%%%%%%%%%%
\subsection{Related work} \label{SS1.1: Related work}
The symmetric nonnegative inverse eigenvalue problem is one specific problem in a wide variety of inverse problems in linear algebra. The surveys \cite{Chu},\cite{Egleston} and more recently \cite{Johnson},\cite{Marijuan},\cite{Lei} give a broad overview of the different inverse eigenvalue subproblems, their applications and the existing strategies to address them. Closely related to the SNIEP are the nonnegative inverse eigenvalue problem (NIEP) for nonnegative matrices and the symmetric doubly stochastic inverse eigenvalue problem (SDIEP) for symmetric nonnegative matrices with row and column sums equal to one. Clearly these problems satisfy NIEP$\supset$SNIEP$\supset$SDIEP, which means that results in either of them can often be translated to the others by appropriate generalization or specialization.
\\
The approach introduced by George Soules \cite{Soules} spawned a promising line of research in the SNIEP. Elsner et al. \cite{Elsner} provided a first important result by describing an explicit construction of all possible Soules vectors. Furthermore, they found that the class of matrices with Soules eigenvectors equals two other classes of matrices, the so-called inverse MMA-matrices and (irreducible) strictly ultrametric matrices; this equivalence was further explored in \cite{Nabben}. Later work described several generalizations of Soules' approach to non-symmetric, non-square matrices \cite{Chen2006},\cite{Chen2007}, reducible matrices \cite{Eubanks} and its application to some problems in linear algebra \cite{Shaked-Monderer},\cite{Catral},\cite{Chen2008}. In \cite{McDonald} and \cite{Loewy}, Soules vectors were used to study the SNIEP for dimensions $N\leq 5$. The wide applicability of Soules' approach was demonstrated recently in \cite{Ellard}, where Ellard and \v{S}migoc proved the equivalence between the eigenvalue conditions that can be derived using Soules' method (allowing for reducible matrices, similar to the proposal of \cite{Eubanks}) and the eigenvalue conditions that follow from a number of other approaches to the SNIEP (see also \cite{Marijuan},\cite{Johnson}). This result of \cite{Ellard} firmly puts Soules' approach among the most fruitful strategies in addressing the SNIEP.
\\
Several other techniques have been developed for tackling the SNIEP, with notable early contributions made by Mirsky and Perfect \cite{Mirsky}, Ciarlet \cite{Ciarlet}, Kellogg \cite{Kellogg} and Fiedler \cite{Fiedler}. Some strategies proceed similarly to Soules' approach and formulate sufficient conditions starting from a specific type of eigenvector matrix, such as the fast Fourier transform matrix \cite{Rojo} and Householder matrices \cite{Zhu}. Other strategies start from the question: ``Given the eigenvalues $\lbrace \lambda_n\rbrace$ of an SNN matrix, which transformations $f$ can be done on these eigenvalues such that the transformed numbers $f(\lbrace \lambda_n\rbrace)$ can still be SNN matrix eigenvalues?" Examples of such transformations were found in \cite{Perfect1955},\cite{Guo},\cite{Smigoc}, and in \cite{Soto2006},\cite{Borobia2008},\cite{Ellard} it was shown how repeated application of these transformations allows to characterize sets of possible eigenvalues. This strategy, together with Soules' approach, leads to the most general sufficient conditions for the SNIEP \cite{Ellard},\cite{Marijuan},\cite{Johnson}. For a more complete overview of the SNIEP and other inverse eigenvalue problems, we again refer to the excellent surveys \cite{Chu},\cite{Egleston},\cite{Johnson},\cite{Marijuan},\cite{Lei}.
%%%%%%%%%%%%%%%%%
%
%
%
%%
%
%
%%%%%%%%%%%%%%%%%%%
%%%%%%%%%%%%%%%%%%%%%%%%%%%%%%%%%%%%%%%%%%%
%
%
%
%
%
%%%%%%%%%%%%%%%%%%%%%%%%%%%%%%%%%%%%%%%%%%%%%
\section{Preliminaries: ordered sets and trees} \label{S2: Preliminaries}
\begin{figure}[h!]
	\centering
    \includegraphics[scale=0.6]{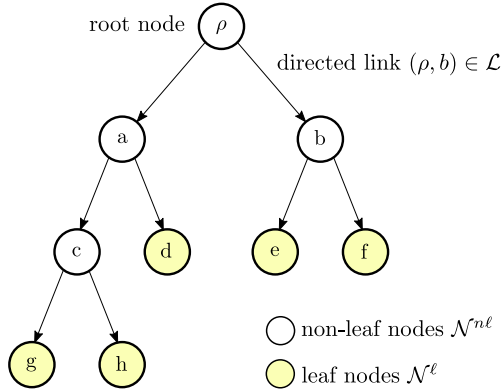}
    \caption{A binary rooted tree $\mathcal{T}$ with leaf nodes $\mathcal{N}^{\ell}=\lbrace d,e,f,g,h\rbrace$, non-leaf nodes $\mathcal{N}^{n\ell}=\lbrace \rho,a,b,c\rbrace$ and root node $\rho$. The partial order $\geqtsub$ orders the non-leaf nodes of $\mathcal{T}$ as $\rho\geqtsub a\geqtsub c$ and $\rho\geqtsub b$ and $(a,b)$ and $(c,b)$ incomparable with respect to $\geqtsub$. This tree will figure as a running example throughout the article.}
    \label{fig: binary rooted tree}
\end{figure}

Before looking further into the inverse eigenvalue problem and Soules' approach, we introduce the relevant mathematical tools and objects. Firstly, we will be interested in ordered sets:
\begin{definition}[Ordered sets]
A \textbf{\textup{totally ordered set}} $(X,\geq)$ and a \textbf{\textup{partially ordered set}} $(X,\geqp)$ consist of a set $X$ of elements which are ordered by a binary relation $``\geq"$ and $``\geqp"$ respectively, that satisfies
\begin{align*}
&x\geq y \textup{~and~} y\geq x \Leftrightarrow x=y, \quad~ x\geq y \textup{~and~} y\geq z \Rightarrow x\geq z,\quad~~ x\geq y \textup{~or~} y\geq x \textup{~or~} x=y
\\
&x\geqp y \textup{~and~} y\geqp x \Leftrightarrow x=y, ~ x\geqp y \textup{~and~} y\geqp z \Rightarrow x\geqp z,~ x\geqp y \textup{~or~} y\geqp x \textup{~or~} x,y
\\
&\hphantom{x\geqp y \textup{~and~} y\geqp x \Leftrightarrow x=y,~x\geqp y \textup{~and~} y\geqp z \Rightarrow x\geqp z,~}\textup{~are incomparable}
\end{align*}
for all triples $x,y,z\in X$.
\end{definition}
The only difference between a total order and a partial order is thus that the latter need not define a relation between all pairs of elements of $X$ while the former does, i.e. a pair of elements can be \emph{incomparable} with respect to a partial order. The consequence of this relaxed requirement is that partially ordered sets can represent more structure in the relations between its elements. For instance, a partial order can capture the reachability structure of any rooted tree (see below), while a total order exclusively corresponds to the reachability of a directed path, clearly a much more constrained structure.
\\
\\
Secondly, we will use the following type of directed graphs:
\begin{definition}[Binary rooted tree]\label{Def0: binary rooted tree}
A binary rooted tree $\mathcal{T}$ is a weakly connected, simple, directed graph which satisfies:
\begin{itemize}[noitemsep]
\item Each node has either
no, or exactly two outgoing links. \textbf{(Binary)}
\item There is one distinct root node $\rho$ such that all links point away from node $\rho$. \textbf{(Rooted)}
\item There is at most one directed path between any pair of nodes. \textbf{(Tree)}
\end{itemize}
\end{definition}
By $\mathcal{N}$ we will denote the set of nodes of $\mathcal{T}$, and by $\mathcal{L}$ the set of directed links, with the convention that a link $(n,m)\in\mathcal{L}$ is directed from $n$ to $m$. A node without outgoing links will be called a \emph{leaf node}, and the set of all leaf nodes is denoted by $\mathcal{N}^{\ell}$. A node $n$ with two outgoing links will be called a \emph{non-leaf node}, and the set of all non-leaf nodes is denoted by $\mathcal{N}^{n\ell} \equiv \mathcal{N} \backslash \mathcal{N}^{\ell}$.
\\
Each binary rooted tree $\mathcal{T}$ has a natural partial order $\geqt$ on the node set $\mathcal{N}$ as
$$
\forall n,m\in\mathcal{N}: ~n \geqt m \text{~if and only if there is a directed path from~}n\text{~to~}m.
$$
As all links in $\mathcal{T}$ point away from the root node $\rho$, we have that $\rho\geqt m$ for all $m\in\mathcal{N}$. If we denote the nodes to which a non-leaf node $n\in\mathcal{N}^{n\ell}$ points by $m^{+}$ and $m^{-}$, i.e. such that $(n,m^{+}),(n,m^{-})\in\mathcal{L}$ holds, then we can define the \emph{leaf descendants} of $n$ via one of its direct descendants as
$$
\mathcal{V}^{+}(n) = \left\lbrace k \in\mathcal{N}^{\ell} ~\bigg\vert~ m^{+} \geqt k  \right\rbrace
\text{~and~}
\mathcal{V}^{-}(n) = \left\lbrace k \in\mathcal{N}^{\ell} ~\bigg\vert~ m^{-} \geqt k  \right\rbrace,
$$
with the union of both sets denoted by $\mathcal{V}(n) = \mathcal{V}^{+}(n)\cup \mathcal{V}^{-}(n)$. Similar to the descendant relations $\mathcal{V}^{\pm}(*)$, we define the following ancestor relation: for some pair of (possibly the same) nodes $n,m\in\mathcal{N}$, their \emph{first common ancestor} is denoted by ${\alpha}_{nm} \triangleq \min\lbrace k\in\mathcal{N}\mid k\geqt m\text{~and~}k\geqt n\rbrace $ with the minimization taken over the partial order $\geqt$. In particular, if $n=m$ then the first ancestor ${\alpha}_{nn}$ simply equals the node $n$. In Figure \ref{fig: binary rooted tree}, an example of a binary rooted tree is given, together with some examples of the introduced notation.
\\
Another property of binary rooted trees is their group of symmetries, called the automorphism group:
\begin{definition}[Automorphism group]
The automorphism group $\Aut(\mathcal{T})$ of a graph $\mathcal{T}$ is the set of all node permutations $\pi$ that preserve the structure of $\mathcal{T}$:
$$
\Aut(\mathcal{T}) = \left\lbrace \pi:\mathcal{N}\rightarrow\mathcal{N} ~\bigg\vert~ (\pi(n),\pi(m))\in\mathcal{L} \Leftrightarrow (n,m)\in\mathcal{L} \right\rbrace
$$
\end{definition}
Importantly, the automorphisms of $\mathcal{T}$ also preserve the partial order; in other words, we have that $\pi(n) \geqt \pi(m) \Leftrightarrow n \geqt m$ for any $\pi\in\Aut(\mathcal{T})$. In some cases, we will consider \emph{weighted} binary rooted trees, where a weight $w(n)$ is assigned to each node $n\in\mathcal{N}$. The automorphism group $\Aut(\mathcal{T}')$ corresponding to such a weighted binary rooted tree $\mathcal{T}'$ then has the additional requirement that any permutation $\pi\in\Aut(\mathcal{T}')$ also needs to preserve the weights of the nodes, i.e. $w(n)=w(\pi(n))$.
%%%%%%%%%%%%%%%
%
%
%
%
%
%
%
\section{Soules bases: from total order to partial order} \label{S3: Soules bases}
\begin{figure}[h!]
\begin{centering}
\includegraphics[scale=0.4]{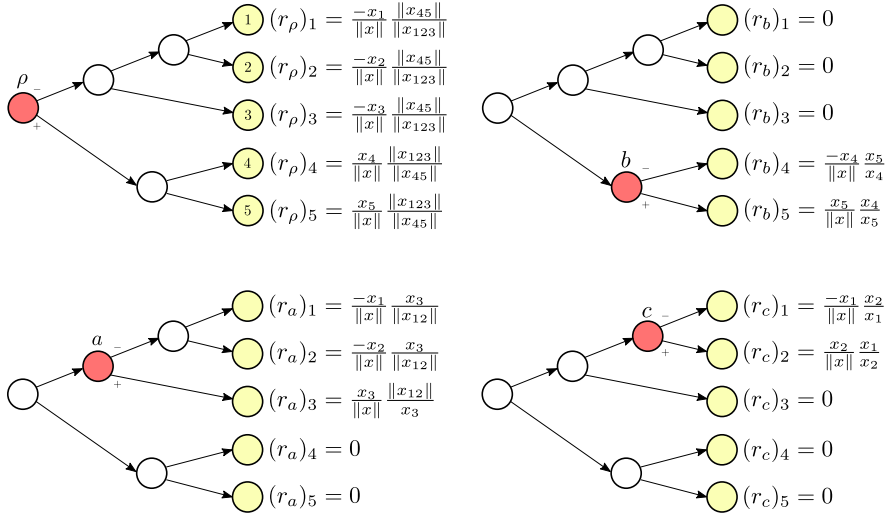}%was scale=0.5
\caption{The Soules vectors $\lbrace r_n\rbrace$ for a binary rooted tree with $N=5$ leaf nodes and positive vector $x$ assigned to the leaf nodes as $x_1,\dots,x_5$; the restricted vectors are abbreviated as $x_{12}\equiv x_{\lbrace 1,2\rbrace},~x_{345}\equiv x_{\lbrace 3,4,5\rbrace},$ etc. For each of the four non-leaf nodes $\lbrace \rho,a,b,c\rbrace$, the Soules vector entries are shown next to the corresponding leaf nodes. Together with the positive vector $r_1=x/\Vert x\Vert$ these four vectors form a Soules basis.}
\label{fig: Soules vectors}
\end{centering}
\end{figure}

As a tool to study the symmetric nonnegative inverse eigenvalue problem, Soules \cite{Soules} constructed a particular set of basis vectors with interesting features for the SNIEP. Elsner, Nabben and Neumann \cite{Elsner} gave the following characterization of these vectors\footnote{In \cite{Elsner}, the definition of a Soules basis does not explicitly consider the Soules vectors as a totally ordered set; however, this total order is implicitly assumed by fixing the Soules vectors as columns of a matrix.}:
\begin{definition}[Totally ordered Soules basis]\label{Def1: TOSB}
A totally ordered set $(\lbrace v_n\rbrace, \geq)$ of orthonormal vectors $v_n\in\mathbb{R}^N$ ordered as $v_1\geq \dots \geq v_N$ with positive vector $v_1$ is called a totally ordered Soules basis, if $S=\sum_{n=1}^N \lambda_n v_n v_n^T$ has nonnegative off-diagonal entries whenever $\lambda_1 \geq \dots \geq \lambda_N$ is satisfied.
\end{definition}
Definition \ref{Def1: TOSB} shows that totally ordered Soules bases (TOSB) are particularly suited to address the SNIEP: as $\min_{i}(S)_{ii}\geq \lambda_N$ holds for the diagonal entries of $S$, we know that if the eigenvalues satisfy $\lambda_1\geq\dots\geq\lambda_N\geq 0$, then matrix $S$ is an SNN matrix. In other words, conditional on the existence of at least one totally ordered Soules basis, we find that any set of non-negative numbers can be the spectrum of a symmetric nonnegative matrix. Existence of totally ordered Soules bases was established by a construction of Soules in \cite{Soules}, who used one particular Soules basis to derive more general sufficient conditions on the eigenvalues of nonnegative matrices. Elsner et al. \cite{Elsner} later gave a complete description of \emph{all possible Soules bases} based on a canonical construction: given a binary rooted tree $\mathcal{T}$ with $N$ leaves and a positive vector $x\in\mathbb{R}^N$, the vectors $\lbrace r_n\rbrace$ are defined as
\begin{equation}\label{eq1: Soules vectors}
r_1 \triangleq \frac{x}{\Vert x\Vert} \textup{~and~} (r_n)_i \triangleq \begin{cases}
\frac{+x_i}{\Vert x_{\mathcal{V}(n)}\Vert} \frac{\Vert x_{\mathcal{V}^{-}(n)}\Vert}{\Vert x_{\mathcal{V}^{+}(n)} \Vert} \textup{~if~}i\in\mathcal{V}^{+}(n)
\\
\frac{-x_i}{\Vert x_{\mathcal{V}(n)}\Vert} \frac{\Vert x_{\mathcal{V}^{+}(n)}\Vert}{\Vert x_{\mathcal{V}^{-}(n)} \Vert} \textup{~if~}i\in\mathcal{V}^{-}(n)
\\
0 \textup{~~otherwise}
\end{cases} \textup{~for all~}n\in\mathcal{N}^{n\ell},
\end{equation}
with $\Vert x_{\mathcal{V}}\Vert^2 = \sum_{i\in\mathcal{V}}x_i^2$ for some subset $\mathcal{V}\subseteq[1,N]$ of entries. Furthermore, the set of vectors $\lbrace r_n\rbrace$ is endowed with a total order $\geq$ which satisfies $n\geqt m\Rightarrow r_n\geq r_m$ for all $n,m\in\mathcal{N}^{n\ell}$ and $r_1\geq r_\rho$ for the root node $\rho$, yielding a totally ordered set of vectors $(\lbrace r_n\rbrace, \geq)$. The construction \eqref{eq1: Soules vectors} thus associates a vector $r_n\in\mathbb{R}^N$ to each non-leaf node $n\in\mathcal{N}^{n\ell}$, and a basis vector of $\mathbb{R}^N$ to each leaf node $i\in\mathcal{N}^{\ell}$, i.e. such that an entry $(*)_i$ of a vector is related\footnote{In fact, the matrix $R=[r_1~\dots~r_N]$ can be seen as a linear map $R:\mathbb{R}^{\vert\mathcal{N}^{n\ell}\vert+1} \rightarrow \mathbb{R}^{\vert\mathcal{N}^{\ell}\vert}$, where non-leaf nodes $n\in\mathcal{N}^{n\ell}$ are basis vectors of the domain of $R$, and leaf nodes $i\in\mathcal{N}^{\ell}$ are basis vectors of the codomain. In this perspective, the entries $(r_n)_i$ of a Soules vector give the coordinates of a basis vector $n\in\mathcal{N}^{n\ell}$ in terms of the basis vectors $i\in\mathcal{N}^{\ell}$ of the codomain.} to leaf node $i$. Figure \ref{fig: Soules vectors} gives an example of construction \eqref{eq1: Soules vectors} for a tree with $N=5$ leaf nodes. A detailed numerical example is given in Section \ref{A2.2: example} (see also Figure \ref{fig: matrix M}).
\\
It can be checked that the set $\lbrace r_n\rbrace$ is a basis for $\mathbb{R}^N$ and, more importantly, that it satisfies the properties of a totally ordered Soules basis:
\begin{theorem}[Elsner, Nabben and Neumann \cite{Elsner}]\label{Th1: TOSB}
~\newline
The ordered set $(\lbrace r_n\rbrace,\geq)$ is a totally ordered Soules basis. Conversely, any totally ordered Soules basis can be constructed from a binary rooted tree as in \eqref{eq1: Soules vectors}.
\end{theorem}
Theorem \ref{Th1: TOSB} thus gives a complete characterization of all totally ordered Soules bases in terms of binary rooted trees, in the strong sense that an ordered set is a totally ordered Soules basis if and only if it is constructed as in \eqref{eq1: Soules vectors} for some $\mathcal{T}$ and $x$. Following this direct connection, we will further refer to the vectors $r_n$ as \emph{Soules vectors}, and the set $\lbrace r_n\rbrace$ as a \emph{Soules basis} (pl: bases).
\\
We now come to \emph{a first central observation} of this article: in the above construction \eqref{eq1: Soules vectors} of the ordered set $(\lbrace r_n\rbrace, \geq)$, the total order imposed on the Soules vectors $\lbrace r_n\rbrace$ is generally not unique. There can be many total orders $\geq$ consistent with the partial order $\geqt$ on the nodes of $\mathcal{T}$, each resulting in a different Soules basis. For instance for the Soules vectors of Figure \ref{fig: Soules vectors}, $(r_\rho\geq r_a \geq r_c \geq r_b)$ and $(r_\rho \geq r_a \geq r_b \geq r_c)$ are two valid but different total orders on the set $\lbrace r_n\rbrace$, resulting in (at least) two different totally ordered Soules bases for a single binary rooted tree. To overcome this issue of redundancy, we suggest to \emph{relax the total order} $r_1\geq \dots \geq r_N$ \emph{to a partial order} $r_n\geqp r_m$.
\begin{definition}[Partially ordered Soules basis]\label{Def2: POSB}
A partially ordered set \sloppy{${(\lbrace v_n\rbrace,\geqp)}$} of orthonormal vectors $v_n\in\mathbb{R}^N$ ordered as $v_1\geq v_n\geqp v_m$ with positive vector $v_1$ is called a partially ordered Soules basis, if $S=\sum \lambda_n v_nv_n^T$ has nonnegative off-diagonal elements whenever $\lambda_1\geq\lambda_n\geqp\lambda_m$ is satisfied.
\end{definition}
By the notation $v_1\geq v_n\geqp v_m$ we denote that there is a partial order $v_n\geqp v_m$ on all vectors, and additionally that the vector $v_1$ precedes all others in the order as $v_1\geq v_n$ for all $n$. Similarly, $\lambda_1\geq\lambda_n\geqp\lambda_m$ means that the real eigenvalues $\lbrace \lambda_n\rbrace$ are ordered consistently with the partial order on the vectors $\lbrace v_n\rbrace$, in other words that $v_n\geqp v_m \Rightarrow\lambda_n\geq\lambda_m$ always holds, and that there is a largest eigenvalue $\lambda_1$.\\
Clearly, the definition of partially ordered Soules bases (POSB) is similar to that of totally ordered Soules bases, apart from the different orders on the set of Soules vectors. This similarity is further exhibited in the following result:
\begin{proposition}\label{Propos1: POSB}
The partially ordered set $(\lbrace r_n\rbrace,\geqt)$ is a partially ordered\linebreak Soules basis. Conversely, any partially ordered Soules basis can be constructed from a binary rooted tree as in \eqref{eq1: Soules vectors}.
\end{proposition}%had to include linebreak to avoid overflow
\emph{Proof:} See Appendix \ref{A1: Proof of POSB}.$\hfill\square\medskip$\\
Proposition \ref{Propos1: POSB} is proven in Appendix \ref{A1: Proof of POSB} using results from Section \ref{SS3.1: Matrices M}, where we further investigate matrices with Soules vectors as eigenvectors. Partially ordered Soules bases thus follow from exactly the same construction as totally ordered Soules bases, but instead of translating the partial order of $\mathcal{T}$ to a total order $\geq$ on the Soules basis $\lbrace r_n\rbrace$, the natural partial order $\geqt$ on the non-leaf nodes of $\mathcal{T}$ is inherited by the vectors $\lbrace r_n\rbrace$ via their association to the leaf nodes of $\mathcal{T}$. For example, following the partial order on the non-leaf nodes $\lbrace \rho,a,b,c\rbrace$ described in Figure \ref{fig: binary rooted tree}, the corresponding Soules vectors in Figure \ref{fig: Soules vectors} satisfy the partial order: $r_\rho\geqt r_a\geqt r_c$ and $r_\rho\geqt r_b$, with $(r_a,r_b)$ and $(r_b,r_c)$ incomparable. Interestingly, the partial order on the Soules vectors is, by construction, equal to the partial order induced by inclusion of the supports of the vectors, i.e. $r_n\geqt r_m \Leftrightarrow \supp(r_n)\supseteq\supp(r_m)$, where $\supp(v)\triangleq\lbrace i\mid (v)_i\neq 0\rbrace$ is the set of non-zero indices of some vector $v$. In principle, any partially ordered or totally ordered Soules basis can thus be represented just by the Soules basis, with the partial order implicit in the supports of the vectors.
\\
\emph{Remark:} Casting the partial order into a total order is not only redundant, it is unnecessarily restrictive: while two eigenvectors $r_n$ and $r_m$, and thus their eigenvalues $\lambda_n$ and $\lambda_m$, can be incomparable with respect to the partial order $\geqt$, the total order $\geq$ on the other hand will always impose a comparability between $r_n$ and $r_m$, such that either $\lambda_n\geq\lambda_m$ or $\lambda_m\geq\lambda_n$ is required. Consequently, the set of eigenvalues $\lbrace \boldsymbol{\lambda}\mid \lambda_1\geq\dots\geq\lambda_N\rbrace$ that satisfy the total order $\geq$, is a subset of the set $\lbrace \boldsymbol{\lambda}\mid\lambda_n\geqt\lambda_m\rbrace$ of eigenvalues that satisfy the partial order (a strict subset, except when the non-leaf nodes of $\mathcal{T}$ form a path).
%%%%%%%%%%%%%%%%%%%%%5
%
%
%
%
%
%%%%%%%%%%%%%%%%%%%%%
\subsection{Matrices with Soules eigenvectors}\label{SS3.1: Matrices M}
In this section, we describe a recursive formula for the entries of matrices $M$ with Soules eigenvectors. Compared to \cite{Elsner} and \cite{Nabben} where the entries $(M)_{ij}$ are studied as well, our proposed formula (Theorem \ref{Th2: M}) highlights how these entries depend on the tree structure $\mathcal{T}$.
\\
\\
We start by introducing the \emph{node function} $s:\mathcal{N}\rightarrow \mathbb{R}$, which assigns a real number to each node of a binary rooted tree $\mathcal{T}$. For a given  positive vector $x$ and a set of values $\lbrace \lambda_n\rbrace$ defined on the non-leaf nodes of $\mathcal{T}$ (i.e. $\lambda_n$ is associated with non-leaf node $n$), we define $s$ recursively as:
\begin{equation}\label{eq2: s node function}
s(\rho) = \frac{\lambda_1-\lambda_\rho}{\Vert x\Vert^2} \text{~and~}
s(m) = 
\begin{dcases}
    s(n) + \frac{\lambda_n-\lambda_m}{\Vert x_{\mathcal{V}(m)}\Vert^2} &\text{~if~} (n,m)\in\mathcal{L}\text{,~and~}m\in\mathcal{N}^{n\ell} \\
    s(n) + \frac{\lambda_n}{\Vert x_{\mathcal{V}(m)}\Vert^2} &\text{~if~}(n,m)\in\mathcal{L}\text{,~and~}m\in\mathcal{N}^{\ell}
\end{dcases}
\end{equation}
where we recall that a link $(n,m)\in\mathcal{L}$ points from the ancestor $n$ to descendant $m$. The values assigned to the nodes of $\mathcal{T}$ are thus defined recursively, starting from the value $s(\rho)$ at the root node, with increments defined over all links $\mathcal{L}$. As these increments depend on the difference between eigenvalues of linked nodes, this suggests to consider the effect of a specified order on the eigenvalues. If we assume that the eigenvalues satisfy the partial order $\geqt$, then we find that $s(m)\geq s(n)$ for all links $(n,m)\in\mathcal{L}$ between non-leaf nodes, which by transitivity yields
\begin{equation}\label{eq2.2: Contravariant order}
\text{If~} \lambda_n\geqt\lambda_m\text{,~then~}n\geqt m\Rightarrow s(n)\leq s(m)\quad\forall n,m\in\mathcal{N}^{n\ell}
\end{equation}
In other words, for partially ordered eigenvalues \emph{the function $s$ obeys the inverse partial order} on $\mathcal{T}$, also called the contravariant order. While defined without explicit reference to Soules bases, the node function $s$ is actually closely related to the entries of matrices with Soules eigenvectors.
\begin{theorem}\label{Th2: M}
The entries of a matrix $M=\sum \lambda_nr_nr_n^T$ with Soules eigenvectors are equal to 
\begin{equation}\label{eq3: M entries}
(M)_{ij} = s(\alpha_{ij})x_ix_j
\end{equation}
\end{theorem}
\emph{Proof:} See Appendix \ref{A2: Proof of M}. $\hfill\square\medskip$\\
For a pair of leaf nodes $i,j\in\mathcal{N}^{\ell}$, we recall that $x_i$ and $x_j$ are entries of the positive vector $x$, that $\alpha_{ij}$ is the first common ancestor and that $\alpha_{ii}$ is equal to the node $i$. A numerical example is given in Section \ref{A2.2: example}, where Theorem \ref{Th2: M} is invoked for a matrix with particular Soules eigenvectors (see also Figure \ref{fig: matrix M}). Expressions \eqref{eq2: s node function} and \eqref{eq3: M entries} that characterize the node function $s$, combined with Theorem \ref{Th2: M} are the crucial ingredients in the proof of Theorem \ref{Th1: TOSB} given in Appendix \ref{A1: Proof of POSB}. Expression \eqref{eq3: M entries} also suggests a concise characterization of the block structure of matrices with Soules eigenvectors: if by $M_{\mathcal{V},\mathcal{S}}$ we denote the submatrix of $M$ with entries in the sets $\mathcal{V},\mathcal{S}\subseteq [1,N]$, and similarly for a vector $x_{\mathcal{V}}$ with entries restricted to $\mathcal{V}$, then by Theorem \ref{Th2: M} we have:
\begin{equation}\label{eq3.2: block structure M}
\forall n\in\mathcal{N}^{n\ell}:~M_{\mathcal{V}^{+}(n),\mathcal{V}^{-}(n)} = s(n) x_{\mathcal{V}^{+}(n)}x_{\mathcal{V}^{-}(n)}^T.
\end{equation}
Expression \eqref{eq3.2: block structure M} states the sub-matrix corresponding to the leaf descendants of some non-leaf node $n$ is a rank-one matrix. All together, this means that $M$ is made up of $N-1$ rank-one blocks on its off-diagonal (one for each non-leaf node). The block structure of matrices with Soules eigenvectors is also investigated in \cite{Nabben} where this structure is referred to as the nested form. 
\\
In the particular case that the eigenvalue $\lambda_1$ corresponding to the positive Soules vector $r_1$ is {strictly} the largest eigenvalue, the definition of the node function \eqref{eq2: s node function} gives that $s(\rho)>0$. If the other eigenvalues are partially ordered, expression \eqref{eq2: s node function} furthermore gives that $s(k)>0$ for all non-leaf nodes $k$. In other words, a unique largest eigenvalue (which is related to irreducibility, see Section \ref{S4: SNIEP}) thus translates by Theorem \ref{Th2: M} into \emph{positivity} of the off-diagonal entries of a matrix with Soules eigenvectors, rather than just the nonnegativity that follows from partially ordered eigenvalues.
\begin{proposition}\label{Propos2: Positive}
A matrix with Soules eigenvectors has \emph{positive} off-diagonal elements whenever its eigenvalues obey the partial order $\lambda_1>\lambda_n\geqt\lambda_m$ corresponding to the partial order on the Soules basis.
\end{proposition}
Finally, we mention that the node function $s$ is an alternative way to parametrize the matrix $M$: from the perspective of the eigendecomposition, this matrix is \linebreak parametrized by the $N$ eigenvalues $\lbrace \lambda_n\rbrace$ and the $N-1$ values of $x$ (as scaling of $x$ does not affect the matrix $M$, this reduces one degree of freedom). From the perspective of the node function $s$, the characterization is given in terms of the $2N-1$ values that $s$ defines on the nodes of $\mathcal{T}$. One representation (eigenvalues and $x$) can be calculated entirely from the other (node function $s$) and thus represents equivalent information.%had to include linebreak between this matrix is *** parametrized
\\
\\
To summarize, we have reformulated the concept of a totally ordered Soules basis (Definition \ref{Def1: TOSB}) into the concept of a partially ordered Soules basis (Definition \ref{Def2: POSB}), which more naturally fits the way these bases are constructed. Next, for matrices $M$ with Soules eigenvectors, expression \eqref{eq2: s node function} and Theorem \ref{Th2: M} combined provide a concise description of the entries $(M)_{ij}$ of these matrices. Expression \eqref{eq2.2: Contravariant order} in particular shows the implications of ordered eigenvalues on the non-negativity of off-diagonal entries, which sheds light on the defining property of Soules vectors.
%%%%%%%%%%%%%%%%%%%%%%%%
%
%
%
%
%%%%%%%%%%%%%%%%%%%%%%%%%
%
%
%
%%%%%%%%%%%%%%%%%%%%%%%%%%%%%

\section{Example}\label{A2.2: example}
\begin{figure}[h!]
    \centering
    \includegraphics[scale=0.6]{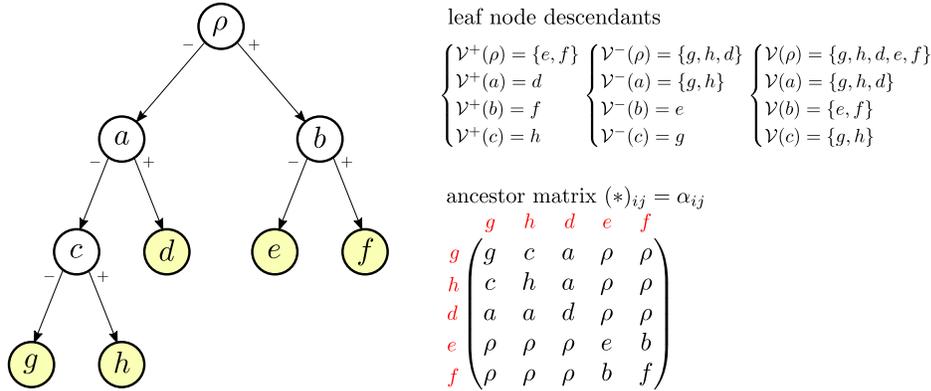}
    \caption{Example of a binary rooted tree $\mathcal{T}$ on $N=5$ leaf nodes. For each non-leaf node, the $\pm$ sign on the outgoing links designates one (out-)neighbour as $m^+$ and the other as $m^-$, e.g. for the root node $\rho$, this indicates that $m^+=b$ and $m^-=a$. The tree partial order $\geqtsub$ then allows to determine the \emph{leaf descendants} $\mathcal{V}^{\pm}(n)$ of each non-leaf node $n$ \textbf{(top)}, and the first common ancestor $\alpha_{ij}$ for each pair of leaf nodes $i$ and $j$ \textbf{(bottom)}.}
    \label{fig: ancestors descendants}
\end{figure}

\begin{figure}[h!]
\begin{centering}
\includegraphics[scale=0.6]{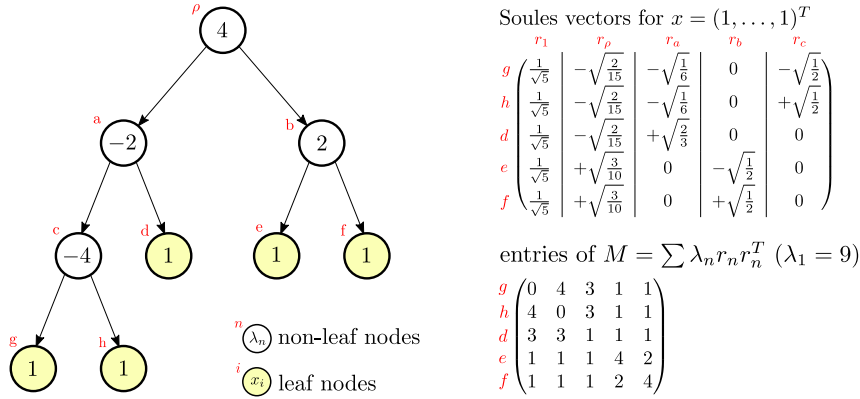}
\caption{\textbf{(top)}~Soules vectors for the binary rooted tree $\mathcal{T}$ on the left. The leaf-node labels indicate the corresponding entry of the positive vector $x$, in this case equal to $x=u$ such that $x_d=\dots=x_f=1$. A quick calculation shows that these vectors are normalized, e.g. $\Vert r_\rho\Vert^2 = 2\tfrac{3}{10}+3\tfrac{2}{15} = 1$, and that pairs of vectors are orthogonal, e.g. $r_\rho^Tr_a = \sqrt{2/15}\left(2\sqrt{1/6}-\sqrt{2/3}\right) = 0$, such that $\lbrace r_n\rbrace$ constitutes a basis for $\mathbb{R}^5$. The Soules vectors for a general positive vector $x$ are shown in Figure \ref{fig: Soules vectors}. \textbf{(bottom)}~Example of a matrix $M=\sum \lambda_nr_nr_n^T$ with the Soules vectors above as eigenvectors. The non-leaf node labels on $\mathcal{T}$ indicate the eigenvalues $\lambda_\rho=4,~\lambda_a=-2,~\lambda_c=-4,~\lambda_b=2$ assigned to these nodes, and the eigenvalue corresponding to $r_1=x/\Vert x\Vert$ (which does not correspond to any non-leaf node) is taken to equal $\lambda_1=9$. As a result of the partially ordered eigenvalues, $M$ has non-negative off-diagonal entries (Definition \ref{Def2: POSB})
}
\label{fig: matrix M}
\end{centering}
\end{figure}

We give a detailed example of the results discussed in Section \ref{S3: Soules bases}. The binary rooted tree $\mathcal{T}$ below has $N=5$ leaf nodes $\mathcal{N}^{\ell}=\lbrace g,h,d,e,f\rbrace$ and four non-leaf nodes $\mathcal{N}^{n\ell}=\lbrace \rho,a,b,c\rbrace$ with root node $\rho$. Following the link directions, the tree partial order $\geqt$ on the non-leaf nodes of $\mathcal{T}$ equals $\rho\geqt a\geqt c$ and $\rho\geqt b$ with $(a,b)$ and $(c,b)$ incomparable. The resulting descendant and ancestor relations between leaf nodes and non-leaf nodes are given in Figure \ref{fig: ancestors descendants}.
To define a Soules basis on $\mathcal{T}$, we choose the positive vector $x=u$, i.e. such that $x_g=x_h=\dots=x_f=1$. This particular choice for $x$ simplifies the norms $\Vert x_{\mathcal{V}^{\pm}(n)}\Vert^2$ to equal the number of nodes in the set $\mathcal{V}^{\pm}(n)$. For instance, we have $\Vert x_{\mathcal{V}(\rho)}\Vert^2 \equiv \vert \mathcal{V}(\rho)\vert = 5$ and $\Vert x_{\mathcal{V}^+(\rho)}\Vert^2 \equiv \vert \mathcal{V}^+(\rho)\vert = 2$. The Soules vectors for $\mathcal{T}$ with $x=u$ are given in Figure \ref{fig: matrix M} following definition \eqref{eq1: Soules vectors}.
\\
Next, we construct a matrix $M=\sum\lambda_nr_nr_n^T$ with these Soules vectors as eigenvectors. We choose the eigenvalues $\lambda_1 = 9, \lambda_\rho = 4,\lambda_a = -2, \lambda_c = -4$ and $\lambda_b=2$, which satisfy the tree partial order. From expression \eqref{eq2: s node function}, we can then calculate the node function $s$ as
$$
\begin{cases}
s(\rho) = \frac{\lambda_1-\lambda_\rho}{\Vert x\Vert^2} = \frac{5}{5}=1
\\
s(a) = s(\rho) + \frac{\lambda_{\rho}-\lambda_a}{\Vert x_{\mathcal{V}(a)}\Vert^2} = 1 + \frac{6}{3} = 3
\\
s(c) = s(a) + \frac{\lambda_a-\lambda_c}{\Vert x_{\mathcal{V}(c)}\Vert^2} = 3 + \frac{2}{2} = 4
\\
s(b) = s(\rho) + \frac{\lambda_{\rho}-\lambda_b}{\Vert x_{\mathcal{V}(b)}\Vert^2} = 1 + \frac{2}{2} = 2
\end{cases}
~
\begin{cases}
s(g)=s(c)+\frac{\lambda_c}{x_g^2} = 4 - \frac{4}{1} = 0\\
s(h)=s(c)+\frac{\lambda_c}{x_h^2} = 4 - \frac{4}{1} = 0\\
s(d)=s(a)+\frac{\lambda_a}{x_d^2} = 3 - \frac{2}{1} = 1\\
s(e)=s(b)+\frac{\lambda_b}{x_e^2} = 2 + \frac{2}{1} = 4\\
s(f)=s(b)+\frac{\lambda_b}{x_f^2} = 2 + \frac{2}{1} = 4
\end{cases}
$$
As a result of the eigenvalues being ordered consistently with $\geqt$, the node function satisfies the inverse partial order on the non-leaf nodes of $\mathcal{T}$, i.e. $s(\rho)<s(a)<s(c)$ and $s(\rho)<s(b)$, as expected by expression \eqref{eq2.2: Contravariant order}. Finally, following Theorem \ref{Th2: M} the entries of $M$ are given by $M=s(\alpha_{ij})x_ix_j=s(\alpha_{ij})$ because $x=u$, with $\alpha_{ij}$ the first common ancestor of leaf nodes $i$ and $j$ (see ancestor matrix in Figure \ref{fig: ancestors descendants}). The resulting matrix is shown in Figure \ref{fig: matrix M}.
\\
%%%%%%%%%%%%
%
%%%%%%%%%%%%
%
%%%%%%%%%%%%
\section{The symmetric nonnegative inverse eigenvalue problem} \label{S4: SNIEP}
The symmetric nonnegative inverse eigenvalue problem (SNIEP) stated in the introduction is concerned with symmetric nonnegative (SNN) matrices $S\in\mathbb{R}^{N\times N}$ which have the following properties:
\begin{align}
&\text{(i)~symmetric:~} S^T = S
\nonumber\\
&\text{(ii)~nonnegative:~} (S)_{ij} \geq 0 \text{~for all~}i,j
\label{eq4: SNN properties}\\
&\text{(iii)~irreducible:~} \nexists \text{~permutation~}P : S = P^T\left(\begin{smallmatrix} A&0\\0&B \end{smallmatrix}\right)P\nonumber
\end{align}
for some symmetric (real) matrices $A,B$. The irreducibility property (iii) is an additional constraint we assume on $S$, since the existence of some $P$ which block-diagonalizes $S$ would mean that the spectrum of $S$ is equal to the union of the spectra of $A$ and $B$, i.e. $\lambda(S)=\lambda(A)\cup\lambda(B)$ where $\lambda(*)$ represents the set of eigenvalues. Thus, if we have a description of the eigenvalue conditions for irreducible matrices, then the conditions for a reducible matrix (of dimension $N\times N$) can be given by the union of the eigenvalue conditions for irreducible matrices of dimensions $N_1,N_2,\dots,N_K$ such that $N = \sum N_i$, taken over all possible compositions $\lbrace N_i\rbrace$. We also remark that zero rows and columns are hereby excluded from $S$, as these would correspond to a block-diagonal form with $A$ or $B$ equal to the all-zero matrix. By assuming irreducibility, we thus focus here on addressing the SNIEP without considering these possible compositions of irreducible matrices to reducible ones.
\\
The SNIEP is concerned with the question of how the SNN properties (i)-(iii) are reflected in the possible eigenvalues of these matrices. In other words, the goal is to describe the \emph{set $\mathcal{S}$ of all possible spectra} for such matrices:
$$
\mathcal{S} \triangleq \left\lbrace \boldsymbol{\lambda}=(\lambda_1,\dots,\lambda_N)^T ~\bigg\vert~ \exists X:X\operatorname{diag}(\boldsymbol{\lambda})X^T\text{~is an SNN matrix}\right\rbrace
$$
with $X$ any possible orthogonal matrix, i.e. satisfying $X^TX = XX^T = I$. A set $\lbrace \lambda_n\rbrace$ such that the vector $\boldsymbol{\lambda}\in\mathcal{S}$ can be realized as the spectrum of an SNN matrix is also called a \emph{realizable set}, and $\boldsymbol{\lambda}$ a \emph{realizable vector}. As any permutation of the vector $\boldsymbol{\lambda}$ corresponds to the same set $\lbrace \lambda_n\rbrace$, all permutations of a realizable vector are also realizable. When describing the realizable vectors as a subset of $\mathbb{R}^N$, it thus makes sense to consider the set $\mathcal{S}$ \emph{without these possible permutations}. Similarly, when a vector $\boldsymbol{\lambda}$ is realizable, the scaled vector $t\boldsymbol{\lambda}$ with $t\in\mathbb{R}_{> 0}$ is realizable as well, simply by scaling the realizing nonnegative matrix by $t$. In principle, it thus suffices to find a description of all realizable vectors \emph{regardless of positive scaling}, which can be done by defining the largest eigenvalue (Perron eigenvalue, see later) to be equal to one.
\subsection{Bounds for the SNIEP: necessary and sufficient conditions}\label{SS4.1: SNIEP bounds}
A first step in addressing the SNIEP follows from translating the SNN properties \eqref{eq4: SNN properties} to necessary features of the eigenvalues. From property (i) that $S$ is symmetric, it follows that the solutions to the eigenequation $Sv_n = \lambda_n v_n$ yield \emph{real eigenvalues} $\lambda_n$ and a set of \emph{orthonormal eigenvectors} $v_n$. For an $N\times N$ matrix $S$, this eigendecomposition can be summarized as
$$
S = \sum_{n=1}^N \lambda_nv_nv_n^T\text{~with~}\lambda_n\in\mathbb{R}\text{~and~}v_n^Tv_m = \mathbbm{1}_{\lbrace n=m\rbrace},
$$
where the indicator function satisfies $\mathbbm{1}_{\lbrace n=m\rbrace}=1$ if $n=m$ and zero otherwise.
\\
From the nonnegative (ii) and irreducible (iii) property, it follows by the Perron-Frobenius theorem that $S$ has a \emph{unique largest eigenvalue} $\lambda_1>0$ which also has the largest absolute value, i.e. $\lambda_1>\vert\lambda_n\vert$ for all $n\neq 1$, with one possible exception: if there is a permutation $P$ such that $S=P^T\left(\begin{smallmatrix}0&A^T\\A&0\end{smallmatrix}\right)P$, then the eigenvalues of $S$ are symmetric around zero and equal to $\pm \vert\lambda(A)\vert$ for the (possibly complex) eigenvalues of $A$, which means that $\lambda_1=-\lambda_N$ is possible. Moreover, the Perron-Frobenius theorem states that the eigenvector $v_1$ corresponding to the unique largest eigenvalue $\lambda_1$ is positive, i.e. $(v_1)_i> 0$ for all $i$. This eigenvalue and eigenvector will further be called the \emph{Perron eigenvalue} and \emph{Perron (eigen)vector}, respectively.
\\
Finally, the nonnegative property (ii) also implies that $\operatorname{tr}(S^t)\geq0$ must hold for any $t\in\mathbb{N}$, and thus that $\sum\lambda_n^t\geq 0$. As these properties (real eigenvalues, unique largest eigenvalue and nonzero sum) hold for any SNN matrix, they are necessary conditions. This information can be summarized by a \emph{set of necessary conditions} $\Omega$ for realizable vectors, defined as
$$
\Omega \triangleq \left\lbrace \boldsymbol{\lambda}\in\mathbb{R}^N ~\bigg\vert~ \lambda_{1} > \lambda_{n\neq 1}\geq -\lambda_1 \text{~and~} \sum_{n=1}^N\lambda_n^t \geq 0 ~~\forall t\in\mathbb{N}\right\rbrace,
$$
which contains the set of all realizable eigenvalue vectors as $\mathcal{S}\subseteq \Omega$. In \cite{McDonald}, a similar description of the necessary conditions is investigated, where (a variant of) $\Omega$ is called the `trace nonnegative polytope'. More subtle necessary conditions have also been identified, see for instance \cite{JohnsonJLL},\cite{LoewyJLL}.
\\
The other way to address the SNIEP is to derive sufficient conditions: these are conditions on the eigenvalues which, if satisfied, guarantee the realizability of a vector (or set). This approach typically relies on considering a specific orthogonal matrix, say $R$, for which the ${N}\choose{2}$ conditions $(R\operatorname{diag}(\boldsymbol{\lambda})R^T)_{ij}\neq 0$ can be reduced to a smaller number of sufficient conditions, say $f_i(\boldsymbol{\lambda})\geq 0$ for $i=1\dots K$ with $K<{{N}\choose{2}}$, tailored to the matrix $R$ specifically. In other words, these approaches generally give solutions of the form ``$f_{i}(\boldsymbol{\lambda})\geq0 \Rightarrow R\operatorname{diag}(\boldsymbol{\lambda})R^T$ is SNN", which then yield a \emph{realizable set of vectors} $\omega(R)$, defined as
$$
\omega(R) \triangleq \left\lbrace \boldsymbol{\lambda}\in\mathbb{R}^N ~\bigg\vert~ f_i(\boldsymbol{\lambda}) \geq 0 \text{~for all~}i \right\rbrace
$$
which is contained in the set of all realizable eigenvalue sequences. All together, the approach to solve the SNIEP by necessary and sufficient conditions thus leads to a characterization of the realizable set $\mathcal{S}$ by upper and lower bounds of the set:
$$
\omega(R) \subseteq \mathcal{S} \subseteq \Omega.
$$
In this article, we consider the SNIEP approach using Soules eigenvectors, i.e. with $R=[r_1~r_2~\dots~r_N]$, and we focus on an investigation of the corresponding realizable set $\omega(R)$. In particular, we show that using the partially ordered Soules basis formulation, the realizable set has a very concise description as a convex cone, with symmetries inherited from the partial order. For other approaches to the SNIEP, we refer to the related work in Section \ref{SS1.1: Related work} and references therein.
%%%%%%%%%%%%%%%%%%%%%%
%
%
%
%%%%%%%%%%%%%%%%%%%%%%5
%
%
%
%%%%%%%%%%%%%%%%%%%%%%
\section{Partially ordered Soules bases and the SNIEP} \label{S5: POSB and SNIEP}
\begin{figure}[h!]
\begin{centering}
\includegraphics[scale=0.35]{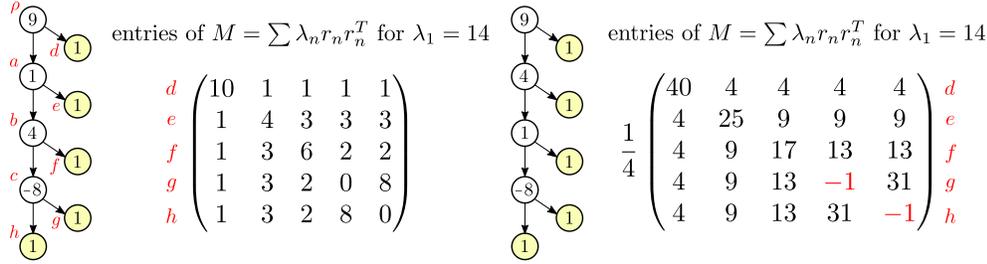} %was scale=0.4
\caption{Example of an assignment of eigenvalues (labels on the non-leaf nodes) which does not satisfy the partial order on the associated tree $\mathcal{T}$, but for which the matrix $M=\sum\lambda_nr_nr_n^T$ is an SNN matrix. In particular, the partial order is violated in the left figure by $\lambda_b=4>\lambda_a=1$ while $a\greattsub b$. Moreover, when these eigenvalues are assigned consistently with the partial order, i.e. with $\lambda_b = 1$ and $\lambda_a=4$ as in the right example, the associated matrix $M$ is no longer an SNN matrix. For the given Soules basis (equivalently, tree $\mathcal{T}$) the set of eigenvalues $\lbrace 14,9,1,4,-8\rbrace$ is thus not realizable according to the partial-order criterion ``$\lambda_1>\lambda_n\geqtsub\lambda_m$ and $s(i)\geq 0$ for all leaves $i$" while it is realizable according to the ``$s(n)\geq0$ for all $n$" criterion.}
\label{fig: counterexample}
\end{centering}
\end{figure}

To study the SNIEP based on partially ordered Soules bases (POSB), we consider matrices with Soules eigenvectors:
$$
M = \sum_{n\in\mathcal{N}^{n\ell}}\lambda_n r_nr_n^T + \lambda_1r_1r_1^T  \text{~with POSB~}(\lbrace r_n\rbrace,\geqt),
$$
with the \emph{strictly} largest eigenvalue $\lambda_1>0$ corresponding to the positive eigenvector $r_1=x/\Vert x\Vert$. Clearly, $M$ satisfies the symmetry property (i) of SNN matrices \eqref{eq4: SNN properties}. By definition of POSBs and Theorem \ref{Th2: M} in particular, we furthermore know that $M$ has positive off-diagonal entries whenever the eigenvalues $\lbrace \lambda_n\rbrace$ obey the partial order $\lambda_1>\lambda_n\geqt\lambda_m$. To satisfy the nonnegativity property (ii), it thus remains to constrain the diagonal of $M$ to be nonnegative as $(M)_{ii}\geq0$. Since $(r_n)_i=0$ for all non-leaf nodes $n$ that are incomparable  (with respect to $\geqt$) to the leaf node $i$, the diagonal requirement translates to $(M)_{ii}=\sum_{n\greatt i} \lambda_n(r_n)_i^2 \geq 0$. Finally, the partially ordered eigenvalues combined with $(M)_{ii}\geq0$ implies that $\lambda_1$ has the largest absolute value\footnote{If we assume that the contrary holds, then there exists a non-leaf node $m$ which is the direct ancestor of some leaf node $i$ (i.e. $(m,i)\in\mathcal{L}$), for which $\lambda_m\leq-\lambda_1$. The corresponding diagonal entry then satisfies $\tfrac{(S)_{ii}}{x_i^2}=\lambda_m(r_m)_i^2 + \sum_{n\greatt m}\lambda_n (r_n)_i^2 +\lambda_1\tfrac{1}{\Vert x\Vert^2} \leq \lambda_m\left({\Vert x_{\mathcal{V}(m)}\Vert^{-2}} - {\Vert x_{\mathcal{V}(m)}\Vert^{-2}} + {\Vert x\Vert^{-2}} - {\Vert x\Vert^{-2}}\right) \leq 0$, using $\lambda_n\leq\lambda_1\leq -\lambda_m$ in the second inequality, which contradicts the nonnegativity of $(M)_{ii}$. Thus, when $(S)_{ii}\geq 0$ for all leaf nodes $i$, this implies by the partial ordering that $\lambda_n\geq -\lambda_1$ for all non-leaf nodes $n$.} such that the irreducibility property (iii) is also satisfied. The sufficient conditions corresponding to a partially ordered Soules basis thus yield the set of realizable vectors
\begin{equation}\label{eq4.2: realizable set POSB}
\omega(\lbrace r_n\rbrace,\geqt) \triangleq \left\lbrace \boldsymbol{\lambda}\in\mathbb{R}^N ~\bigg\vert~ \lambda_1>\lambda_n\geqt\lambda_m\text{~and~} \sum\nolimits_{n\greatt i}\lambda_n(r_n)_i^2\geq0\text{~for all~}i\in\mathcal{N}^{\ell}\right\rbrace,
\end{equation}
which is a subset of all realizable eigenvalue vectors $\omega(\lbrace r_n\rbrace, \geqt)\subseteq\mathcal{S}$. We will further call $\lambda_1>\lambda_n \geqt \lambda_m$ the \emph{order inequalities} and  $\sum\lambda_n(r_n)_i^2\geq 0$ the \emph{diagonal inequalities}.
Using the node function $s$, we can also write the diagonal inequalities as $s(i)\geq 0$ for leaf nodes $i$, where the dependence of $s$ on the eigenvalues $\boldsymbol{\lambda}$ is implicit. This formulation illustrates that the realizable set $\omega(\lbrace r_n\rbrace,\geqt)$ fundamentally depends on the binary rooted tree $\mathcal{T}$. Indeed \emph{the weighted binary rooted tree $\mathcal{T}_x$} where each node $n$ has weight $\Vert x_{\mathcal{V}(n)}\Vert$, contains all information\footnote{As all non-leaf node weights depend on the leaf-node weights as $\Vert x_{\mathcal{V}(n)}\Vert$, in principle the leaf node weights carry all necessary weight information.} to encode the order inequalities and the node function $s$. An alternative description for the realizable set is thus
\begin{equation}\label{eq5: realizable set Tx}
\omega(\mathcal{T}_x) = \left\lbrace \boldsymbol{\lambda}\in\mathbb{R}^N~\bigg\vert~ \lambda_1>\lambda_n\geqt\lambda_m \text{~and~}s(i)\geq0 ~\forall i\in\mathcal{N}^{\ell}\right\rbrace\equiv \omega(\lbrace r_n\rbrace,\geqt)
\end{equation}
for any binary rooted tree with positively weighted leaves $\mathcal{T}_x$, which is agnostic to the specific construction of Soules vectors $\lbrace r_n\rbrace$. We stress that the description of the realizable sets \eqref{eq4.2: realizable set POSB} and \eqref{eq5: realizable set Tx} depends on the definition of partially ordered Soules bases (Definition \ref{Def2: POSB}), which is newly introduced in this article. Appendix \ref{SS5.2: Comparison} discusses how the SNIEP sufficient conditions based on partially ordered Soules bases differ from the conditions based on totally ordered Soules bases.
\\
\\
\emph{Remark:} In deriving the realizable sets \eqref{eq4.2: realizable set POSB},\eqref{eq5: realizable set Tx}, we use the Soules property (i.e. ordered eigenvalues) to determine realizable vectors for which $M$ is an SNN matrix. This approach that originally taken by Soules in \cite{Soules} and is followed in later works \cite{McDonald},\cite{Loewy},\cite{Ellard} that study the SNIEP using Soules vectors. However, partially ordered eigenvalues are \emph{not necessary} for off-diagonal entries of $M$ to be nonnegative. Figure \ref{fig: counterexample} shows an example of eigenvalues $\lbrace \lambda_n\rbrace$ which do not satisfy the partial order $\geqt$, but for which $s(n)\geq 0$ for all nodes $n\in\mathcal{N}$. Even more, the example shows that there exist eigenvalues which are realizable based on the requirements $s(n)\geq 0$ for $n\in\mathcal{N}$ but which are not realizable according to the conditions $\lambda_1>\lambda_n\geqt\lambda_m$ and $s(i)\geq 0$ for $i\in\mathcal{N}^{\ell}$ (while the opposite can never happen). The example in figure \ref{fig: counterexample} implies that a more complete description of the realizable eigenvalues for a given Soules basis $\lbrace r_n\rbrace$ should be based on the conditions that $s(n)\geq 0$, rather than the ordered-eigenvalues approach. We do not further explore the implications of this enhanced realizability criterion in this article, but it might be a promising line of future research.
\subsection{Properties of the realizable set $\omega(\mathcal{T}_x)$}\label{SS5.1: Properties of set}
For a binary tree with $N$ leaves (and thus $N-1$ non-leaf nodes), the realizable set $\omega(\mathcal{T}_x)$ set is determined by $2N-1$ inequalities in total: $N$ diagonal inequalities $s(i)\geq 0$ (one for each non-leaf nodes), $1$ strict order inequality $\lambda_1>\lambda_\rho$ and $N-2$ order inequalities $\lambda_n\geq \lambda_m$ (one for each of the links) from which all other order inequalities follow by transitivity. 
\\
A first observation from expression \eqref{eq4.2: realizable set POSB} is that \emph{the order and diagonal inequalities are linear} in the eigenvalues: both $\lambda_1>\lambda_n\geqt\lambda_m$ and $\sum \lambda_n(r_n)_i^2\geq 0$ are linear in $\boldsymbol{\lambda}$. Geometrically speaking, each of these linear inequalities determines a \emph{half-space} in $\mathbb{R}^N$ in which the eigenvalues $\boldsymbol{\lambda}$ ``live" that satisfy this inequality (see for instance \cite[Ch. 2]{Dattorro}). We define the half-spaces $\mathcal{H}_{n,m} = \lbrace\boldsymbol{\lambda}\in\mathbb{R}^N\mid\lambda_n-\lambda_m\geq0\rbrace$ for non-leaf nodes $n,m$, and $\mathcal{H}_{i,i}=\lbrace\boldsymbol{\lambda}\in\mathbb{R}^N\mid s(i)\geq 0\rbrace$ for leaf nodes $i$. Moreover, since all inequalities are of the form $b^T\boldsymbol{\lambda}\geq0$ (with some coefficient vector $b$), these half-spaces go through the origin $0\in\mathbb{R}^N$. Given that a realizable vector $\boldsymbol{\lambda}$ must satisfy each inequality at the same time, the realizable set equals the intersection of these $2N-1$ half-spaces:
$$
\omega(\mathcal{T}_x) = \bigcap_{\substack{(n,m)\in\mathcal{L}^+,\\n=m=1\dots N}} \mathcal{H}_{n,m},
$$
where $\mathcal{L}^+=\mathcal{L}\cup(1,\rho)$, such that the open half-space determined by the strict inequality $\lambda_1>\lambda_\rho$ is included. As the intersection of (open or closed) half-spaces that share a point on their boundary determines a convex cone, we thus arrive at the following characterization:
\begin{property}[Cone]\label{Proper1: Cone}
The realizable set $\omega(\mathcal{T}_x)$ is a convex cone: for $v,w\in\omega(\mathcal{T}_x)$ and $t_1,t_2\in\mathbb{R}_{\geq0}$ the conic combination $t_1 v+t_2 w$ is also in $\omega(\mathcal{T}_x)$.
\end{property}
If we consider the realizable set modulo positive scaling, e.g. by fixing $\lambda_1=1$, then the realizable set $\omega(\mathcal{T}_x)\cap\lbrace\boldsymbol{\lambda}\mid\lambda_1=1\rbrace$ is a \emph{conic section}, which is a convex set.
\\
\\
A second observation follows from expression \eqref{eq5: realizable set Tx} for the realizable set based on properties of $\mathcal{T}_x$ alone: if a permutation of the node set, say $\pi$, leaves the structure of $\mathcal{T}_x$ unchanged, i.e. $(n,m)\in\mathcal{L}\Leftrightarrow (\pi(n),\pi(m))\in\mathcal{L}$ and $\Vert x_{\mathcal{V}(n)}\Vert = \Vert x_{\mathcal{V}(\pi(n))}\Vert$ for all nodes $n,m$, then the order and diagonal inequalities for $\boldsymbol{\lambda}$ are the same when $\lambda_n$ is assigned to node $n$, as when $\lambda_{\pi(n)}$ is assigned to node $n$, denoted by the permuted vector $\boldsymbol{\lambda}^{\pi}$. In other words, permutations that are automorphisms of the weighted tree $\mathcal{T}_x$ also determine permutations that leave the realizability of a vector $\boldsymbol{\lambda}$ invariant\footnote{A permutation $\pi\in\Aut(\mathcal{T}_x)$ is defined on all nodes of $\mathcal{T}_x$ as $\pi:\mathcal{N}\rightarrow\mathcal{N}$. When applied to an eigenvalue vector $\boldsymbol{\lambda}$, only the permutations of the non-leaf nodes are taken into account, i.e. $\lambda_n \rightarrow \lambda_{\pi(n)}$ for all $n\in\mathcal{N}^{n\ell}$. Moreover, as $r_1$ has no corresponding node in $\mathcal{T}_x$, we define any permutation $\pi\in\Aut(\mathcal{T}_x)$ to leave the assignment of $\lambda_1$ unchanged, i.e. $\lambda_{\pi(1)} = \lambda_1$.}. As a consequence, we have the following property:
\begin{property}[Symmetry]\label{Proper2: Symmetry}
The realizable set $\omega(\mathcal{T}_x)$ is symmetric with respect to permutations from the automorphism group of $\mathcal{T}_x$: for $v\in\omega(\mathcal{T}_x)$ and $\pi\in\operatorname{Aut}(\mathcal{T}_x)$, the vector $v^{\pi}$ is also in $\omega(\mathcal{T}_x)$.
\end{property}
\emph{Proof:} See Appendix \ref{A3: Proof of symmetry}.$\hfill\square\medskip$\\
Whenever the positive vector is a constant vector $x=u$ with $u=(1,\dots,1)^T$, the automorphism group of $\mathcal{T}_u$ is equal to the automorphism group of the unweighted tree $\mathcal{T}$, whose size is at least as large as the automorphism group for any non-constant weights. The case $x=u$ coincides with a subproblem of the SNIEP: an SNN matrix with a constant Perron vector has constant row and column sums and thus corresponds to a symmetric doubly stochastic matrix. The relevant inverse eigenvalue problem for the $x=u$ case is thus the SDIEP (see Section \ref{SS1.1: Related work}).
\\
\\
Finally, it is revealing to combine the cone and symmetry properties: for any realizable vector $v$ and positive scalars $\theta_\pi$, the vector $\sum_{\pi\in \Aut(\mathcal{T}_x)} \theta_\pi v^{\pi}$ is also realizable. Now, if $P^{\pi}$ denotes the permutation matrix such that $v^{\pi}=P^{\pi}v$, then we can define the cone of $N\times N$ matrices generated by the automorphisms of $\mathcal{T}_x$ as
$$
\mathscr{P}(\mathcal{T}_x) = \bigg\lbrace \sum_{\pi\in\Aut(\mathcal{T}_x)}\theta_\pi P^{\pi}\text{~with~}\theta_{\pi}\in\mathbb{R}_{\geq 0}\bigg\rbrace
$$
which we call \emph{the permutation cone of $\Aut(\mathcal{T}_x)$}. The combination of properties \ref{Proper1: Cone} and \ref{Proper2: Symmetry} is then compactly characterized as follows:
\begin{property}\label{Proper3: Permutation cone}
The realizable set $\omega(\mathcal{T}_x)$ is closed under multiplication by matrices from the permutation cone of $\Aut(\mathcal{T}_x)$: for $v\in\omega(\mathcal{T}_x)$ and $P\in\mathscr{P}(\mathcal{T}_x)$ the vector $Pv$ is also in $\omega(\mathcal{T}_x)$.
\end{property}
Similarly, if we consider the realizable set modulo positive scaling, say $\omega(\mathcal{T}_x)\vert_{\lambda_1=1}$, we find that for all realizable vectors $v$ and matrices $P\in\widetilde{\mathscr{P}}(\mathcal{T}_x) = \big\lbrace \sum_{\pi\in\Aut(\mathcal{T}_x)} \theta_\pi P^\pi$ with $\sum\theta_\pi =1\big\rbrace$, the vector $Pv$ is again realizable. A convex set of the form $\widetilde{\mathscr{P}}(\mathcal{T}_x)$ is commonly known \cite{Baumeister} as the \emph{permutation polytope of the group $\Aut(\mathcal{T}_x)$}. Property \ref{Proper3: Permutation cone} might provide an interesting perspective on the realizable set, as permutation polytopes are well-studied objects for various groups \cite{Baumeister}.
%%%%%[ref weird break of the line with \mathscr{P}]
\\
\\
To summarize, we discuss how to approach the SNIEP based on necessary and sufficient conditions. We derive a set of realizable vectors $\omega(\mathcal{T}_x)$ for SNN matrices by using Soules vectors as eigenvectors. Moreover, we show that this set is a convex cone with symmetries inherited from the tree automorphism group $\Aut(\mathcal{T}_x)$. Importantly, the results in this section are obtained by employing partially ordered Soules bases. In Appendix \ref{SS5.2: Comparison}, we show that a derivation of sufficient conditions based on totally ordered Soules bases (as is usually the case, e.g. in \cite{McDonald},\cite{Loewy},\cite{Ellard}) fails to capture the properties of the realizable set derived from POSBs.
%%%%%%%%%%%%%%%%%
%
%
%%%%%%%%%%%%%%%%%
%
%
%%%%%%%%%%%%%%%%%%%
\section{Constructing Laplacian matrices with a given spectrum} \label{S7: Laplacian}
In this section, we describe a number of applications of Soules vectors to graph theory and network science. We show how a graph Laplacian matrix can be constructed with Soules eigenvectors and how this allows to realize \emph{any} positive spectrum. Furthermore, we discuss how the structure of the associated tree $\mathcal{T}$ is reflected in the properties of the Laplacian matrix and its corresponding graph.
\subsection{The Laplacian matrix}
An undirected weighted graph $G$ consists of a set of nodes $\mathcal{N}_G$ and a set of undirected links $\mathcal{L}_G$, where each link $(i,j)\in\mathcal{L}_G$ has an associated \emph{positive} weight $w_{ij}$. The \emph{Laplacian matrix} $Q$ of a weighted graph on $N$ nodes is a symmetric $N\times N$ matrix with entries
$$
(Q)_{ij} = 
\begin{cases}
d_i &\text{~if~}i=j
\\
-w_{ij} &\text{~if~}(i,j)\in\mathcal{L}_G
\\
0 &\text{~otherwise}
\end{cases}
$$
where $d_i = \sum w_{ij}$ is the weighted degree of a node $i$, equal to the sum of the weights of all links connected to $i$. For a connected graph, the Laplacian matrix is irreducible and has a single zero eigenvalue \cite[art. 80]{pvm_GS} with corresponding eigenvector equal to the constant vector $u/\sqrt{N}$. Furthermore, as the Laplacian is a positive semi-definite matrix, this zero eigenvalue is strictly the smallest eigenvalue.
\\
The algebraic representation of a graphs structure by its Laplacian matrix is central in algebraic and spectral graph theory and many relations are known between (the spectrum of) the Laplacian and properties of the graph, see \cite{Chung},\cite{pvm_GS},\cite{Mohar} for an overview of these results.
\\
As all link weights are necessarily positive, the Laplacian matrix $Q$ has non-positive off-diagonal entries. Moreover, since $Q$ has a zero eigenvalue corresponding to the positive eigenvector $u/\sqrt{N}$, we find the following Soules construction of Laplacian matrices:
\begin{proposition} \label{Propos3: Laplacian}
For a partially ordered Soules basis $(\lbrace r_n\rbrace,\geqt)$ with constant positive vector $x=u$, the matrix $Q = \sum \mu_nr_nr_n^T$ is the Laplacian matrix of an undirected, weighted graph on $N$ nodes whenever the eigenvalue corresponding to $r_1=u/\sqrt{N}$ equals $\mu_1=0$, and the other eigenvalues satisfy the inverse partial order $n\geqt m\Rightarrow \mu_n\leq \mu_m$.
\end{proposition}
\emph{Proof:} The matrix $-Q$ has Soules eigenvectors and eigenvalues $\lambda_k=-\mu_k$ that satisfy the partial order $\lambda_1\geq \lambda_n\geqt\lambda_m$. By Definition \ref{Def2: POSB} of Soules vectors, it thus follows that $(Q)_{ij}\leq 0$. Then, as $Qu=0$, the diagonal of the Laplacian equals $(Q)_{ii} = -\sum_{j\neq i}(Q)_{ij}$, which means that the entries of $Q$ satisfy the definition of a Laplacian matrix. $\hfill\square\medskip$\\
A numerical example of Proposition \ref{Propos3: Laplacian} is given in Figure \ref{fig: Laplacian and EEPs}\color{black}.The most interesting feature of our proposed construction of Laplacian matrices is that \emph{any (positive) spectrum can be realized}\footnote{There are many ways to assign a given set of eigenvalues to the non-leaf nodes of a binary rooted tree. The number of ways in which a total order can be embedded into any non-isomorphic binary rooted tree on $N$ leaf nodes is calculated in \cite{Murtagh} and equals \emph{Sequence A000111} in the On-line Encyclopedia of Integer Sequences. As shown in \cite[p. 296]{Flajolet}, this sequence grows as $2N!(2/\pi)^N$ for $N\rightarrow \infty$, which means there can be as many different Laplacian matrices realizing a given set of eigenvalues}. The possibility to construct Laplacian matrices with tunable eigenvalues and eigenvectors (by choice of $\mathcal{T}$) might prove a useful tool for the study of graphs and networks. This utility is illustrated by the work of Forrow et al. \cite{Forrow}, where a Laplacian matrix with a predescribed spectrum (discovered independently of the theory of Soules vectors) is used to study the effect of band-gapped eigenvalues in a variety of physical systems. In fact, the theory of Soules vectors and in particular Proposition \ref{Propos3: Laplacian} can be invoked directly to generalize the work in \cite{Forrow} by enabling the construction of not just one, but a whole range of Laplacian matrices with a given spectrum. In \cite{Clauset}, Clauset, Moore and Newman propose a hierarchical random graph model that successfully models real-world network properties; this random graph model can be interpreted as choosing a particular function $s$ on a binary rooted tree $\mathcal{T}$ and is thus closely related to the theory of Soules vectors.
%%%%
%
%%%%
\subsection{Effective resistance}
\label{SS7.1: Spectral bounds}
In a positively weighted graph $G$, the \emph{effective resistance} $\omega_{ij}$ is a distance function between pairs of nodes $i,j$. While originally defined in context of electrical circuit theory, the effective resistance has been related to a wide range of concepts in graph theory and network science such as random walks \cite{Doyle}, distance functions on graphs \cite{Klein}, graph embeddings \cite{Krl_simplex}, node centrality \cite{Krl}, sparsification \cite{Spielman} and many others. This ubiquity of the effective resistance in graph theory can be understood partly by its close relation to the graph Laplacian $Q$; the effective resistance $\omega_{ij}$ between a pair of nodes $i$ and $j$ is equal to
\begin{equation}\label{eq7: Effective resistance}
\omega_{ij} = (Q^{\dagger})_{ii}+(Q^{\dagger})_{jj}-2(Q^{\dagger})_{ij} = \sum_{n=2}^N\mu_n^{-1}\left[(v_n)_i - (v_n)_j\right]^2,
\end{equation}
where $Q^{\dagger}$ is the pseudoinverse of $Q$, and $\lbrace v_n\rbrace$ are the eigenvectors and $\lbrace \mu_n\rbrace$ the non-zero eigenvalues for a general Laplacian matrix \cite{Krl}. In the particular case that the Laplacian matrix $Q$ has Soules eigenvectors $v_n\equiv r_n$, we find that the effective resistance can be related to the binary rooted tree $\mathcal{T}$ as follows:
\begin{theorem}\label{thm4: Effective resistance}
In a graph $G$ whose Laplacian matrix has Soules eigenvectors with binary rooted tree $\mathcal{T}$, the effective resistance between a pair of nodes equals the shortest (undirected) weighted path-length between the corresponding leaf nodes in $\mathcal{T}$:
$$
\omega_{ij} = \sum_{(n,m)\in\mathcal{P}(i,j)}w_{nm}
$$
where $\mathcal{P}(i,j)=\lbrace (i,n_1),\dots,(n_K,j)\rbrace$ contains the links in the shortest undirected path between leaf nodes $i$ and $j$, and with link weights $w_{nm}=\frac{\mu_n^{-1}-\mu_m^{-1}}{\vert\mathcal{V}(m)\vert}$ if $m$ is a non-leaf node and $w_{nm}=\mu_{m}^{-1}$ if $m$ is a leaf node. 
\end{theorem}
\emph{Proof:} See Appendix \ref{A5: proof of spectral bounds}. $\hfill\square\medskip$\\
Theorem \ref{thm4: Effective resistance} states that the nodes of $G$ can be embedded into the leaves of $\mathcal{T}$ (in some sense, the boundary of $\mathcal{T}$) such that the effective resistance metric $\omega_{ij}$ on the graph is preserved as the shortest path metric on the tree. 
%%%%
%
%%%%
\subsection{External equitable partitions}
\begin{figure}[h!]
    \centering
    \includegraphics[scale=0.5]{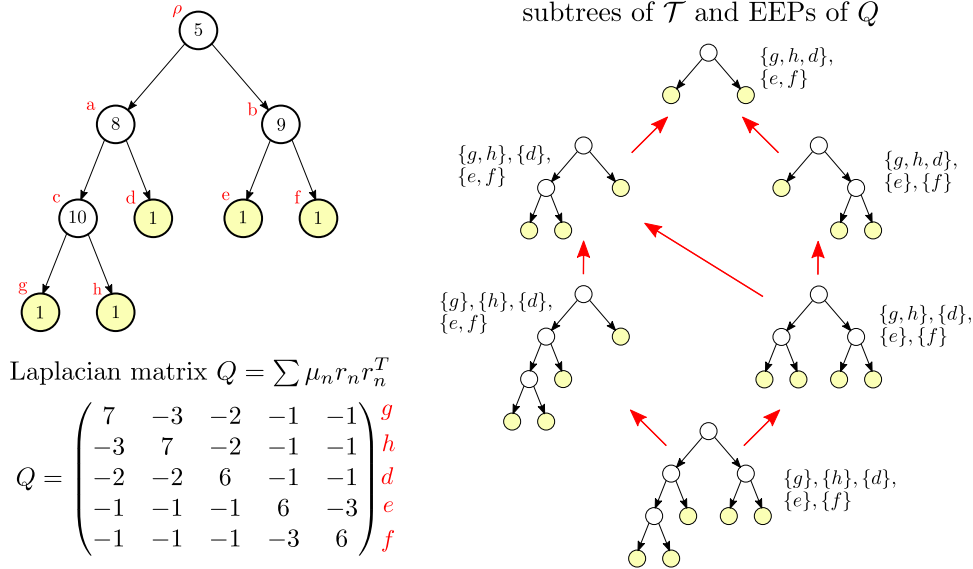}
    \caption{\textbf{(left)} Numerical example of a $5\times 5$ Laplacian matrix with Soules eigenvectors. The leaf-node labels indicate the corresponding entries of the positive vector $x=u$. The non-leaf node labels indicate the corresponding eigenvalues $\mu_\rho = 5, \mu_a=8,\mu_c=10, \mu_b=9$ and $\mu_1=0$ corresponding to the constant eigenvector $r_1=u/\sqrt{N}$.~\textbf{(right)} The subtrees $\widehat{\mathcal{T}}$ (except the trivial subtree $\widehat{\mathcal{T}}=\rho$) of $\mathcal{T}$ are given, with arrows indicating the partial order, i.e. $``\color{red}\rightarrow\color{black}"\equiv``\supseteq"$. Next to each subtree, the corresponding equitable partition (EEP) of $Q$ is given.}
    \label{fig: Laplacian and EEPs}
\end{figure}

An external equitable partition (EEP) of a graph $G$ consists of a partitioning of the nodes of $G$ into disjoint subsets
$\mathcal{N}_1\cup\mathcal{N}_2\cup\dots\cup\mathcal{N}_K=\mathcal{N}_G$ such that all nodes in a set $\mathcal{N}_k$ have \emph{the same number of links} to nodes in the other sets $\mathcal{N}_{m\neq k}$. In the case of weighted graphs, the total weight of links to nodes in the other sets needs to be the same. In terms of the (weighted) Laplacian matrix $Q$, this translates to:
$$
\text{~A partition~}\lbrace \mathcal{N}_k\rbrace_{k=1}^K\text{~is an EEP if~}\sum_{j\in\mathcal{N}_{m}}(Q)_{i_1j} = \sum_{j\in\mathcal{N}_m}(Q)_{i_2j}\text{~for all~}i_1,i_2\in\mathcal{N}_{k\neq m}.
$$
In other words, an external equitable partition divides a graph into sets of nodes $\mathcal{N}_k$ with a similar connectivity pattern to the rest of the network. Interestingly, the existence of an EEP can lead to \emph{synchronization between nodes} when a dynamical process takes place on the graph. This is the case, for instance, for consensus dynamics \cite{OClery}, epidemic spreading\footnote{In the case of epidemics of networks, a stronger form of similarity between nodes is required: the number of links between nodes of the same partition also needs to be the same, i.e. $\sum_{j\in\mathcal{N}_k}(Q)_{i_1j}=\sum_{j\in\mathcal{N}_k}(Q)_{i_2j}$ for all $i_1,i_2\in\mathcal{N}_k$.} \cite{krl_UMFF},\cite{Bonaccorsi} and coupled oscillators \cite{Schaub}.
\\
When graphs are constructed from Soules vectors as in Proposition \ref{Propos3: Laplacian}, the particular block structure of the Laplacian matrix $Q$ means that \emph{many different external equitable partitions} can be found. We remark that the nodes of a graph $G$ are related to the leaves of the binary tree $\mathcal{T}$ such that a partition of the nodes can also be represented by a partition of the leaves. By $\widehat{\mathcal{T}}\subseteq\mathcal{T}$ we denote a subtree of $\mathcal{T}$, i.e. a binary rooted tree with links $\widehat{\mathcal{L}}\subseteq\mathcal{L}$ and nodes $\widehat{\mathcal{N}}\subseteq\mathcal{N}$ (and leaf nodes $\widehat{\mathcal{N}}^\ell$). The equitable partitions of $Q$ can then be characterized as follows:
\begin{proposition}\label{Propos6: equitable partitions}
For any subtree $\widehat{\mathcal{T}}\subseteq\mathcal{T}$ with $K$ leaf nodes $\widehat{\mathcal{N}}^\ell$, the partitioning of the leaf nodes $\mathcal{N}^\ell$ of $\mathcal{T}$ into $K$ partitions defined as
$$
\mathcal{N}_k = \left\lbrace i\in\mathcal{N}^\ell ~\bigg\vert~ i \leqt k\right\rbrace \textup{~for all~}k\in\widehat{\mathcal{N}}^\ell
$$
determines an external equitable partition for the nodes of any graph whose Laplacian matrix has Soules eigenvectors based on $\mathcal{T}$.
\end{proposition}
\textbf{Proof:} We consider a fixed tree $\mathcal{T}$ and subtree $\widehat{\mathcal{T}}\subseteq\mathcal{T}$. Since $Q$ is a matrix with Soules eigenvectors, its entries are given by Theorem \ref{Th2: M} as $(Q)_{ij} = s(\alpha_{ij})x_ix_j=s(\alpha_{ij})$ since $x_i=x_j=1$ and with $\alpha_{ij}$ the first common ancestor of leaf nodes $i,j\in\mathcal{N}^\ell$. Each of the partitions $\mathcal{N}_k$ is associated with a leaf node $k\in\widehat{\mathcal{N}}^\ell$ of $\widehat{\mathcal{T}}$. Hence, the definition of external equitable partitions requires that for any $k\neq m\in\widehat{\mathcal{N}}^\ell$, the equality $\sum_{j\in\mathcal{N}_m}(Q)_{i_1j}=\sum_{j\in\mathcal{N}_m}(Q)_{i_2j}$ is satisfied for all $i_1,i_2\in\mathcal{N}_k$. Introducing the proposed partitions in this expression allows the two sides of this requirement to be written as
\begin{equation}\label{eq7.2: proof of EEP}
\begin{dcases}
\sum\limits_{j\in\mathcal{N}_m}(Q)_{i_1j} = \sum\limits_{j\leqt m}(Q)_{i_1j} = \sum\limits_{j\leqt m}s(\alpha_{i_1j})=\sum\limits_{j\leqt m}s(\alpha_{km}) = s(\alpha_{km})\vert\mathcal{V}(m)\vert\\
\sum\limits_{j\in\mathcal{N}_m}(Q)_{i_2j} = \sum\limits_{j\leqt m}(Q)_{i_2j} = \sum\limits_{j\leqt m}s(\alpha_{i_2j})=\sum\limits_{j\leqt m}s(\alpha_{km}) = s(\alpha_{km})\vert\mathcal{V}(m)\vert\\
\end{dcases}
\end{equation}
where we used the fact that $\alpha_{ij} = \alpha_{nm}$ if $i\in\mathcal{V}(n),j\in\mathcal{V}(m)$ and $n,m$ incomparable with respect to $\geqt$, which translates to $\alpha_{i_1j}=\alpha_{km}$ and $\alpha_{i_2j}=\alpha_{km}$ since $k$ and $m$ are leaves in $\widehat{\mathcal{T}}$ and are thus incomparable with respect to $\geqt$. From \eqref{eq7.2: proof of EEP} follows that the sets $\mathcal{N}_k$ satisfy the definition of an EEP which completes the proof.$\hfill\square\medskip$\\
Proposition \eqref{Propos6: equitable partitions} states that weighted graphs constructed using Soules eigenvectors admit equitable partitions\footnote{In fact, the proposed partitions satisfy a stronger property: $(Q)_{i_1j_1}=(Q)_{i_2j_2}$ for all $i_1,i_2\in\mathcal{N}_k$ and $j_1,j_2\in\mathcal{N}_m$ and all $k\neq m$.} based on subtrees $\widehat{\mathcal{T}}$ of the corresponding tree $\mathcal{T}$. The set of all subtrees can be seen as a partially ordered set $(\lbrace \mathcal{T}_i\mid\mathcal{T}_i\subseteq\mathcal{T}\rbrace,\geqp)$, with the partial order defined by the subtree relation as $\mathcal{T}_i\geqp\mathcal{T}_j$ if and only if $\mathcal{T}_j\subseteq\mathcal{T}_i$. This partial order also reflects how the equitable partitions for different subtrees are related. If by $\lbrace*\rbrace\overset{\text{ref}}{\subseteq}\lbrace*\rbrace$ we denote that one partition (the first) \emph{refines} another partition (the second), i.e.
$$
\lbrace \mathcal{N}_k^{(j)}\rbrace \overset{\text{ref}}{\subseteq}\lbrace \mathcal{N}_k^{(i)}\rbrace\Longleftrightarrow
\bigcup_{k=1}^{K^{(j)}} \mathcal{N}_k^{(j)} = \bigcup_{k=1}^{K^{(i)}}\mathcal{N}_k^{(i)}\text{~and~}\forall k,\exists m:  \mathcal{N}_k^{(i)}\subseteq\mathcal{N}^{(j)}_m,
$$
then we have that $\mathcal{T}_i\geqp\mathcal{T}_j$ implies $\lbrace \mathcal{N}_k^{(i)}\rbrace \overset{\text{ref}}{\subseteq}\lbrace \mathcal{N}_k^{(j)}\rbrace$. In other words, the partial order on the subtrees describes a \emph{hierarchy of equitable partitions} where ascending in the partial order corresponds to refining the partition and descending to coarsening it.In graphs with Soules eigenvectors, EEPs exist across all scales, corresponding to the whole range of subtrees between $\widehat{\mathcal{T}}=\mathcal{T}$ and $\widehat{\mathcal{T}}=\rho$, yielding the finest and coarsest equitable partitions respectively. From the perspective of dynamical processes taking place on the graph $G$, this means that the structure of $G$ supports synchronization at different scales for processes governed by the Laplacian. The relation between subtrees of $\mathcal{T}$ and equitable partitions of $Q$ is illustrated in Figure \ref{fig: Laplacian and EEPs}.
%%%%%%%%%%%%%%%
%
%
%
%%%%%%%%%%%%%%%%%%
%
%
%
%%%%%%%%%%%%%%%%
\section{Conclusion} \label{S8: Conclusion}
This article discusses various aspects of Soules vectors, which were introduced in \cite{Soules} and \cite{Elsner}. We give a self-contained description of the construction of these Soules bases $\lbrace r_n\rbrace$ based on binary rooted trees $\mathcal{T}$, and discuss how matrices $M=\sum \lambda_n r_nr_n^T$ with Soules eigenvectors can be used in context of the SNIEP. Moreover, we introduce a new application of Soules vectors: constructing graph Laplacian matrices with any desired positive spectrum.
\\
The first contribution of this article is Definition \ref{Def2: POSB}, where we introduce partially ordered Soules bases as a relaxation of the commonly used totally ordered Soules bases. Our proposed definition more naturally captures the defining \emph{Soules property}: if the eigenvalues of $M$ are ordered consistently with the partial order on the eigenvectors then $M$ has nonnegative off-diagonal entries.
\\
A second result is related to the entries of matrices with Soules eigenvectors. We introduce a function $s:\mathcal{N}\rightarrow\mathbb{R}$ on the nodes of $\mathcal{T}$ and show in Theorem \ref{Th2: M} that each entry of $M$ corresponds to $s(n)$ for some node $n$. Moreover, we find that the node function $s$ obeys the inverse partial order compared to the eigenvalues on the non-leaf nodes of $\mathcal{T}$ (Expression \eqref{eq2.2: Contravariant order}), which gives an intuitive explanation of the Soules property.
\\
Thirdly, in context of the SNIEP, we consider the set of realizable eigenvalues for matrices with Soules eigenvectors, i.e. the set $\omega(\mathcal{T}_x)$ which consists of the eigenvalues for which $M$ is a symmetric nonnegative matrix. The main contribution here is the characterization of $\omega(\mathcal{T}_x)$ as a conic set with symmetries corresponding to the automorphism group of $\mathcal{T}_x$. Stated differently, we find that the realizable set is closed under multiplication by matrices from the permutation cone of $\Aut(\mathcal{T}_x)$. These results are stated in Property \ref{Proper1: Cone}--\ref{Proper3: Permutation cone}.
\\
Finally, we describe how to construct graph Laplacian matrices with Soules eigenvectors (Proposition \ref{Propos3: Laplacian}), which allows to generate (weighted) graphs $G$ with any given Laplacian spectrum. We furthermore discuss how, by construction, the structure of these graphs $G$ is related to the associated tree $\mathcal{T}$. In particular, we show that \emph{equitable partitions} of $G$ can be found from subtrees of $\mathcal{T}$ and that the effective resistance between nodes of $G$ corresponds to the shortest path-length on (a weighted variant of) $\mathcal{T}$. This work opens interesting lines of research in graph inference, for instance, with the possibility to infer the most likely tree of a real-world network, and thus to provide a coarse-grained representation in terms of its hierarchical structure.
\section*{Acknowledgements}
Karel Devriendt was supported by The Alan Turing Institute under the EPSRC grant EP/N510129/1.

%%%%%%%%%
%APPENDIX
%%%%%%%%%
\appendix
\section*{Appendix}
\section{Proof of Proposition \ref{Propos1: POSB} (Partially ordered Soules basis)} \label{A1: Proof of POSB}
We consider the matrix $M = \sum\lambda_n r_nr_n^T$ with Soules eigenvectors $\lbrace r_n\rbrace$ and eigenvalues $\lbrace \lambda_n\rbrace$. By Theorem \ref{Th2: M}, this matrix has entries equal to $(M)_{ij}=x_ix_js(\alpha_{ij})$, where $\alpha_{ij}$ is the first common ancestor of leaf nodes $i$ and $j$, and with the node function $s$ given by \eqref{eq2: s node function}.
\\
We first show that the partially ordered set $(\lbrace r_n\rbrace,\geqt)$ \emph{satisfies the definition} of a partially ordered Soules basis (Definition \ref{Def2: POSB}). If the eigenvalues of $M$ satisfy the partial order $\lambda_1\geq\lambda_n\geqt\lambda_m$ consistent with the order of the eigenvectors $\lbrace r_n\rbrace$, then the node function satisfies $s(\rho)\geq 0$ and it obeys the inverse partial order \eqref{eq2.2: Contravariant order}. By transitivity, this means that $s(n)\geq 0$ for all non-leaf nodes $n$. By Theorem \ref{Th2: M} and the positivity of $x$, we then have that $(M)_{ij}\geq 0$ for all off-diagonal entries $i\neq j$ as a result of the eigenvalues being partially ordered. This proves that $(\lbrace r_n\rbrace,\geqt)$ satisfies the definition of a partially ordered Soules basis.
\\
Next, we \emph{prove the converse statement}: any partially ordered Soules basis can be constructed from a binary tree as in \eqref{eq1: Soules vectors}.\emph{~Assume that the opposite is true}, then there exists a partially ordered set $(\lbrace v_n\rbrace,\geqp)$ of vectors $\lbrace v_n\rbrace$ which are not constructible from a binary rooted tree, and which has the following property: for any set of eigenvalues $\lbrace \mu_n\rbrace$ that satisfies the partial order $\geqp$, the matrix $M = \sum\mu_nv_nv_n^T$ has nonnegative off-diagonal entries. Now, for a particular set of eigenvalues $\lbrace \mu_n^{\star}\rbrace$ which satisfy the partial order as $\mu_1^{\star}\geq\mu_n^{\star}\geqp\mu_m^{\star}$, we define a total order $\geq$ on the set $\lbrace v_n\rbrace$ induced by ordering the eigenvalues according to decreasing values, i.e. $\mu_n^{\star}\geq\mu_m^{\star}\Rightarrow v_n\geq v_m$. By construction, this totally ordered set $(\lbrace v_n\rbrace,\geq)$ now has the property that $S=\sum\lambda_kv_nv_n^T$ has nonnegative off-diagonal entries whenever the eigenvalues $\lbrace \lambda_n\rbrace$ satisfy the total order $\geq$. This property is equal to Definition \ref{Def1: TOSB}, which means that the ordered set $(\lbrace v_n\rbrace, \geq)$ is a Soules basis. By the converse of Theorem \ref{Th1: TOSB} the set of vectors $\lbrace v_n\rbrace$ is thus always constructible as in \eqref{eq1: Soules vectors} from a binary rooted tree, \emph{which contradicts the assumption} and thus proves the converse of Proposition \ref{Propos1: POSB}.~$\square$
%%%%%%%%%%%%%%%%%%%%%%%%
%
%
%
%%%%%%%%%%%%%%%%%%%%%%%%%%%%%
%
%
%
%%%%%%%%%%%%%%%%%%%%%%%%%%%%%%
\section{Proof of Theorem \ref{Th2: M} (entries of $M$)} \label{A2: Proof of M}
The off-diagonal entries of a matrix $M$ with Soules eigenvectors are given by
\begin{equation}\label{eqA1: Matrix with Soules eigenvectors}
(M)_{ij} = \sum_{n\in\mathcal{N}^{n\ell}}\lambda_n(r_n)_i(r_n)_j + \lambda_1(r_1)_i(r_1)_j\quad\forall i\neq j.
\end{equation}
By definition \eqref{eq1: Soules vectors} of the Soules vectors, we have that $(r_n)_i \neq 0$ only if $n\greatt i$ and similarly, that $(r_n)_j\neq 0$ only if $n\greatt j$. Consequently, we have that $(r_n)_i(r_n)_j\neq 0$ only if $n\geqt \alpha_{ij}$ with $\alpha_{ij}$ the first common ancestor of leaf nodes $i$ and $j$. Expression \eqref{eqA1: Matrix with Soules eigenvectors} can then be written as
\begin{equation}\label{eqA2: Matrix with Soules eigenvectors 2}
(M)_{ij} = \sum_{n \geqt \alpha_{ij}} \lambda_n(r_n)_i(r_n)_j + \lambda_1\frac{x_ix_j}{\Vert x\Vert^2}.
\end{equation}
In order to retrieve the recursion of the node function $s$, we consider a pair of linked non-leaf nodes $(n,m)\in\mathcal{L}$ (pointing from $n$ to $m$) and corresponding entries $(M)_{ij}$ and $(M)_{\beta\gamma}$ such that $n$ and $m$ are their corresponding first common ancestors, i.e. $m\equiv \alpha_{ij}$ and $n\equiv \alpha_{\beta\gamma}$. Using expression \eqref{eqA2: Matrix with Soules eigenvectors 2}, we can then write the difference $(M)_{ij}-(M)_{\beta\gamma}$ as
\begin{align}
(M)_{ij}-(M)_{\beta\gamma} &= \sum_{k \geqt m} \lambda_k(r_k)_i (r_k)_j - \sum_{k\geqt n} \lambda_k(r_k)_\beta (r_k)_\gamma + \frac{\lambda_1}{\Vert x\Vert^2}(x_ix_j-x_\beta x_\gamma)\nonumber
\\
&= \lambda_m(r_m)_i(r_m)_j + \lambda_n (r_n)_i(r_n)_j - \lambda_n (r_n)_\beta(r_n)_\gamma \nonumber
\\
&\hphantom{=}+ \sum_{k \greatt n} \lambda_k\left[(r_k)_i(r_k)_j-(r_k)_\beta(r_k)_\gamma\right] + \frac{\lambda_1}{\Vert x\Vert^2}(x_ix_j-x_\beta x_\gamma)\label{eqA3: Matrix with Soules eigenvectors difference}
\end{align}
As the leaf nodes $i,j,\beta,\gamma$ are all in $\mathcal{V}(n)$, we have that $\frac{(r_k)_i(r_k)_j}{x_ix_j} = \frac{(r_k)_\beta(r_k)_\gamma}{x_\beta x_\gamma}$ for all $k \greatt n$. If we then consider the difference between $(M)_{ij}$ and $(M)_{\beta\gamma}$ normalized by $x_ix_j$ and $x_\beta x_\gamma$ respectively, we find that the fourth term and fifth term of \eqref{eqA3: Matrix with Soules eigenvectors difference} are zero. Further introducing the values of $r_k$, we then find
\begin{align}
\frac{(M)_{ij}}{x_ix_j} - \frac{(M)_{\beta\gamma}}{x_\beta x_\gamma} &= \frac{-\lambda_m}{\Vert x_{\mathcal{V}(m)}\Vert^2} + \lambda_n\frac{\Vert x_{\mathcal{V}(n)}\Vert^2 - \Vert x_{\mathcal{V}(m)}\Vert^2}{\Vert x_{\mathcal{V}(m)}\Vert^2\Vert x_{\mathcal{V}(n)}\Vert^2} + \frac{\lambda_n}{\Vert x_{\mathcal{V}(n)}\Vert^2}\nonumber
\\ 
&= \frac{\lambda_n-\lambda_m}{\Vert x_{\mathcal{V}(m)}\Vert^2}\label{eqA3: Matrix with Soules eigenvectors difference 2}
\end{align}
which is valid for any pair of linked non-leaf nodes $n$ and $m$.
For a pair of leaf nodes $i$ and $j$ with the root node $\rho$ as first common ancestor, i.e. $\alpha_{ij}\equiv \rho$, we obtain the following expression for $(M)_{ij}$:
$$
(M)_{ij} = \lambda_\rho(r_\rho)_i(r_\rho)_j + \lambda_1\frac{x_ix_j}{\Vert x\Vert^2}.
$$
Introducing the values for $r_\rho$ and normalizing by $x_ix_j$, we find that
\begin{equation}\label{eqA4: Matrix with Soules eigenvectors difference 3}
\frac{(M)_{ij}}{x_ix_j} = \frac{\lambda_1-\lambda_\rho}{\Vert x\Vert^2}.
\end{equation}
Finally, for the diagonal entries $(M)_{ii}$ we define $n$ to be the direct ancestor of the leaf-node $i$, in other words $(n,i)\in\mathcal{L}$, which then yields
$$
(M)_{ii} = \sum_{k \geqt n} \lambda_k(r_k)_i^2
$$
Next, we define $j$ to be any other leaf node for which $\alpha_{ij}\equiv n$. The difference between the diagonal entry $(M)_{ii}$ and the off-diagonal entry $(M)_{ij}$ is then given by
\begin{align}
(M)_{ii}-(M)_{ij} &= \sum_{k \geqt n} 
\lambda_k\left[ (r_k)_i^2-(r_k)_i(r_k)_j\right]\nonumber
\\
&= \lambda_n\left[(r_n)_i^2-(r_n)_i(r_n)_j\right] + \sum_{k \greatt n} \lambda_k\left[(r_k)_i^2 - (r_k)_i(r_k)_j\right].\label{eqA5: Matrix with Soules eigenvectors difference 4}
\end{align}
For non-leaf nodes $k \greatt \alpha_{ij}$ we have that $\tfrac{(r_k)_i}{x_i}=\tfrac{(r_k)_j}{x_j}$ and thus that $\tfrac{(r_k)_i^2}{x_i^2}-\tfrac{(r_k)_i(r_k)_j}{x_ix_j}=0$. Considering the difference between the normalized entries of $M$ thus yields
\begin{align}
\frac{(M)_{ii}}{x_i^2} - \frac{(M)_{ij}}{x_ix_j} 
&= \lambda_n\left[\frac{\Vert x_{\mathcal{V}(n)} \Vert^2-\Vert x_{\mathcal{V}(i)}\Vert^2}{\Vert x_{\mathcal{V}(i)}\Vert^2.\Vert x_{\mathcal{V}(n)}\Vert^2} + \frac{1}{\Vert x_{\mathcal{V}(n)}\Vert^2}\right]\nonumber
\\
&= \frac{\lambda_n}{\Vert x_{\mathcal{V}(i)}\Vert^2}.\label{eqA6: difference}
\end{align}
Finally, we see that if we define $s(n) = \tfrac{(M)_{ij}}{x_ix_j}$ for any non-leaf nodes $i$ and $j$ such that $\alpha_{ij}=n$, then \eqref{eqA3: Matrix with Soules eigenvectors difference 2}, \eqref{eqA4: Matrix with Soules eigenvectors difference 3} and \eqref{eqA6: difference} give precisely the definition of the node function $s$ as in \eqref{eq2: s node function}. $\hfill\square$
%%%%%%%%%%%%
%
%
%%%%%%%%%%%%
%
%
%%%%%%%%%%%%
\section{Proof of Property \ref{Proper2: Symmetry} (Symmetry of realizable set)}\label{A3: Proof of symmetry}
Given a realizable vector $\boldsymbol{\lambda}$ which assigns eigenvalues $\lambda_n$ to non-leaf nodes $n\in\mathcal{N}^{n\ell}$, we consider the permuted vector $\boldsymbol{\lambda}^{\pi}$ (with $\lambda_{\pi(1)}=\lambda_1$) which assigns eigenvalues $\lambda_{\pi(n)}$ to non-leaf nodes $n\in\mathcal{N}^{n\ell}$ for a permutation 
\begin{equation}\label{eqA6.2: Automorphisms}
\pi\in\Aut(\mathcal{T}_x) \Rightarrow (n,m)\in\mathcal{L}\Leftrightarrow (\pi(n),\pi(m))\in\mathcal{L}\text{,~and~}\Vert x_{\mathcal{V}(n)}\Vert = \Vert x_{\mathcal{V}(\pi(n))}\Vert,
\end{equation}
corresponding to automorphisms of the weighted binary rooted tree $\mathcal{T}_x$. We emphasize that an automorphism $\pi$ is defined for all nodes of the tree $\mathcal{T}_x$ (i.e. including the leaf nodes), but that by applying this permutation to an eigenvalue vector as $\boldsymbol{\lambda}^\pi$ we only consider the permutations of the non-leaf nodes.
\\
For the \emph{order inequalities}, we have that 
$\lambda_1>\lambda_n\geqt\lambda_m$ by \eqref{eqA6.2: Automorphisms} implies that also $\lambda_{\pi(1)}>\lambda_{\pi(n)}\geqt\lambda_{\pi(m)}$ also holds, which means that all permuted vectors $\boldsymbol{\lambda}^{\pi}$ satisfy the order inequalities. For the \emph{diagonal inequalities}, we consider the node function $s$ based on $\boldsymbol{\lambda}$ and the node function $s'$ based on the permuted eigenvalues $\boldsymbol{\lambda}^{\pi}$, which are defined as
\begin{equation}\label{eqA7: s and s'}
\begin{cases}
s(\rho) = \tfrac{\lambda_1-\lambda_\rho}{\Vert x\Vert^2}\\
s(m) = s(n)+\tfrac{\lambda_n-\lambda_m}{\Vert x_{\mathcal{V}(m)}\Vert^2}\\
s(i) = s(k)+\tfrac{\lambda_{k}}{\Vert x_{\mathcal{V}(i)}\Vert^2}
\end{cases}
\quad
\begin{cases}
s'(\rho) = \tfrac{\lambda_{1}-\lambda_{\rho}}{\Vert x\Vert^2}\\
s'(m) = s'(n)+\tfrac{\lambda_{\pi(n)}-\lambda_{\pi(m)}}{\Vert x_{\mathcal{V}(m)}\Vert^2}\\
s'(i) = s'(k)+\tfrac{\lambda_{\pi(k)}}{\Vert x_{\mathcal{V}(i)}\Vert^2},
\end{cases}
\end{equation}
where $(n,m)\in\mathcal{L}$ and $(k,i)\in\mathcal{L}$ for non-leaf nodes $n,m$ and leaf node $i$. Invoking the automorphism properties \eqref{eqA6.2: Automorphisms} in the definition of $s'$ in \eqref{eqA7: s and s'} shows that $s'(m) = s(\pi(m))$ for all nodes. Hence, if all diagonal inequalities $s(i)\geq 0$ are satisfied for $\boldsymbol{\lambda}$, then the diagonal inequalities $s'(i)\geq 0$ are also satisfied for the permuted vector $\boldsymbol{\lambda}^\pi$.$\hfill\square\medskip$
%%%%%%%%%%%%%%%
%
%
%
%%%%%%%%%%%%%%%%
\section{Comparison with totally ordered Soules bases}\label{SS5.2: Comparison} Section \ref{S5: POSB and SNIEP} describes how sufficient conditions for the symmetric nonnegative inverse eigenvalue can be found by considering matrices with partially ordered Soules eigenvectors. In Section \ref{SS5.1: Properties of set}, the resulting sufficient conditions are shown to have a number of interesting properties that allow to characterize the realizable set $\omega(\mathcal{T}_x)$ as a convex set with certain symmetries. 
\\
Here, we discuss what happens when instead of POSBs, a totally ordered Soules basis is used to derive sufficient conditions. Following the same derivations as in Section \ref{S5: POSB and SNIEP}, a totally ordered Soules basis $(\lbrace r_n\rbrace, \geq)$ has a corresponding set of realizable vectors
\begin{equation}\label{eq6: realizable set TOSB}
\left\lbrace \boldsymbol{\lambda}\in\mathbb{R}^N ~\bigg\vert~ \lambda_{1}> \dots \geq \lambda_{N} \text{~and~} \sum\nolimits_{n\greatt i} (r_n)_i^2 \lambda_{n}\geq 0\text{~for all~} i\in\mathcal{N}^{\ell} \right\rbrace,
\end{equation}
As discussed before - and apparent in expression \eqref{eq6: realizable set TOSB} - the total order on the eigenvalues $\lbrace \lambda_n\rbrace$ is unnecessarily restrictive. For instance, if a pair of nodes $n,m$ is incomparable with respect to $\geqt$, they can appear in total orders where $\lambda_n\geq\lambda_m$ is a condition and others where $\lambda_m\geq \lambda_n$ is the condition, either of which restricts the possible values of $\lambda_n$ and $\lambda_m$. Consequently, the set \eqref{eq6: realizable set TOSB} does not fully capture the potential of Soules vectors to describe sufficient conditions for the SNIEP\footnote{An exception is the case where all non-leaf nodes of $\mathcal{T}$ are arranged in a path, i.e. when all but one non-leaf nodes have one leaf descendant and one non-leaf descendant. In that case, which was in fact the basis constructed by Soules in \cite{Soules}, the partial order on the Soules vectors equals the total order and the realizable sets \eqref{eq4.2: realizable set POSB} and \eqref{eq6: realizable set TOSB} coincide.}.
\\
To overcome the unnecessary restrictions of imposing a total order, one should combine all possible total orders when defining the realizable set. If we use a bijection $\gamma:[2,N]\rightarrow \mathcal{N}^{n\ell}$ to denote the arrangement of the eigenvalues $\lbrace \lambda_n\rbrace$ in a total order, i.e. such that $\lambda_{1}\geq\lambda_{\gamma(2)}\geq\dots\geq\lambda_{\gamma(N)}$, then we can describe \emph{all total orders $\geq$
compatible with a partial order $\geqt$} by the set
$$
\Gamma(\mathcal{T}) = \left\lbrace \gamma:[2,N]\rightarrow \mathcal{N}^{n\ell} ~\bigg\vert~ t_1\leq t_2 \Leftarrow \gamma(t_1) \geqt \gamma(t_2)\text{~for all~} t_1,t_2\in[2,N]\right\rbrace.
$$
This set of permutations $\Gamma(\mathcal{T})$ allows to compactly formulate the realizable set $\omega(\lbrace r_n\rbrace)$ for a Soules basis with all possible total orders, as:
\begin{equation}\label{eq6.2: realizable set TOSB, Union}
\omega(\lbrace r_n\rbrace,\geq) =\bigcup_{\gamma\in\Gamma(\mathcal{T})}  \left\lbrace \boldsymbol{\lambda}\in\mathbb{R}^N ~\bigg\vert~ \lambda_{1}>\dots \geq \lambda_{\gamma(N)} \text{~and~} \sum (r_n)_i^2 \lambda_{n}\geq 0, \forall i\in\mathcal{N}^{\ell} \right\rbrace.
\end{equation}
We call \eqref{eq6.2: realizable set TOSB, Union} the totally ordered Soules basis (TOSB) expression of the realizable set of $\lbrace r_n\rbrace$ and \eqref{eq5: realizable set Tx} the partially ordered Soules basis (POSB) expression. The TOSB expression and POSB expression of the realizable set both determine the same set, i.e. $\omega(\lbrace r_n\rbrace,\geq) =\omega(\lbrace \mathcal{T}_x\rbrace)$, however, properties \ref{Proper1: Cone}-\ref{Proper3: Permutation cone} are (arguably) much more apparent in the POSB expression \eqref{eq4.2: realizable set POSB} as compared to the TOSB expression \eqref{eq6.2: realizable set TOSB, Union}. Firstly, while \eqref{eq6: realizable set TOSB} is a convex cone for each total order individually, the union over different convex cones, as in \eqref{eq6.2: realizable set TOSB, Union}, is generally not a cone nor even a convex set. Secondly, the problem of testing whether a vector $\boldsymbol{\lambda}$ is in $\omega(\lbrace r_n\rbrace,\geqt)$ amounts to testing whether all of the $2N-1$ inequalities hold, while testing whether it is in $\omega(\lbrace r_n\rbrace,\geq)$ amounts to $2N-1$ tests in the best case, and $\vert \Gamma(\mathcal{T})\vert(2N-1)$ tests in the worst case. As $\vert\Gamma(\mathcal{T})\vert\geq 1$ with equality if and only if the non-leaf nodes of $\mathcal{T}$ form a path, it is more efficient\footnote{Testing membership of the realizable set $\omega(\lbrace r_n\rbrace,\geqt)$ can be done in $O(N)$ time, but this only holds for a fixed tree Soules basis $\lbrace r_n\rbrace$. In contrast, Borobia and Canogar \cite{BorobiaNP} have recently shown that testing the general realizability, including the case of irreducible matrices, is NP-hard} to test whether a vector is realizable based on the POSB expression \eqref{eq5: realizable set Tx} than based on the TOSB expression \eqref{eq6: realizable set TOSB}.
%%%%%%%%%%%%%%%%%%
%
%
%
%%%%%%%%%%%%%%%%%%
%
%
%
%%%%%%%%%%%%%%%%%%
%%%%%%%%%%%%%%%%%%%%%%5
%
%
%%%%%%%%%%%%%%%%%%%%%%
%
%
%%%%%%%%%%%%%%%%%%%%%%%%%
%%%%%%%%%%%%%%%%5
%
%
%%%%%%%%%%%%%%%%%
%
%
%%%%%%%%%%%%%%%%
\section{Proof of Theorem \ref{thm4: Effective resistance} (Effective resistance)}\label{A5: proof of spectral bounds}
Translating the spectral form \eqref{eq7: Effective resistance} of the effective resistance to the case of Laplacian matrices with Soules eigenvectors, and using the fact that $(r_n)_i=(r_n)_j$ and thus $((r_n)_i-(r_n)_j)^2=0$ whenever $n\greatt \alpha_{ij}$, we obtain the expression
\begin{align}\label{eqA9: Reduced effective resistance}
\omega_{ij} &= \sum_{n\leqt \alpha_{ij}}\mu_n^{-1}((r_k)_i-(r_k)_j)^2 
\end{align}
For the quadratic terms, we have that $(r_n)^2_i\neq 0$ if and only if $n\greatt i$, similarly for $(r_n)^2_j$, and for the cross terms that $(r_n)_i(r_n)_j \neq 0$ if and only if $n\geqt \alpha_{ij}$ (by definition \eqref{eq1: Soules vectors}). Expression \eqref{eqA9: Reduced effective resistance} for the effective resistance thus becomes
\begin{equation}\label{eqA10: Reduced effective resistance 2}
\omega_{ij} = \sum_{i \lesst n \leqt \alpha_{ij}} \mu_n^{-1}(r_n)_i^2 + \sum_{j \lesst n \leqt \alpha_{ij}} \mu_n^{-1}(r_n)_j^2 + \mu_{\alpha_{ij}}^{-1}\frac{2}{\Vert x_{\mathcal{V}(\alpha_{ij})}\Vert^2}.
\end{equation}
If we denote the path of nodes between $i$ and $\alpha_{ij}$ by $(k_0\lesst k_1\lesst \dots\lesst k_{\ell})$ such that $k_0=i$, $k_{\ell}=a_{ij}$ and $(k_{l},k_{l-1})\in\mathcal{L}$, then we can write $(r_{k_l})_i^2 = \frac{\vert\mathcal{V}(k_l)\vert-\vert\mathcal{V}(k_{l-1})\vert}{\vert\mathcal{V}(k_l)\vert\vert\mathcal{V}(k_{l-1})\vert}
$. Similarly, denoting the path between $j$ and $\alpha_{ij}$ by $(m_0\lesst m_1\lesst\dots\lesst m_{\ell'})$, we can write the effective resistance as
\begin{equation*}
    \omega_{ij} = \sum_{l=1}^\ell\mu_{k_l}^{-1}\left(\frac{1}{\vert\mathcal{V}(k_{l-1})\vert}-\frac{1}{\vert\mathcal{V}(k_l)\vert}\right) + \sum_{l=1}^{\ell'}\mu_{m_l}^{-1}\left(\frac{1}{\vert\mathcal{V}(m_{l-1})\vert}-\frac{1}{\vert\mathcal{V}(m_l)\vert}\right) + \frac{2\mu_{\alpha_{ij}}^{-1}}{\vert\mathcal{V}(\alpha_{ij})\vert}.
\end{equation*}
This expression can be written as a sum over the links of $\mathcal{T}$ as
\begin{equation}\label{eqA11: node-based expression}
    \omega_{ij} = \mu_i^{-1} +  \sum_{l=2}^\ell\frac{\mu_{k_l}^{-1}-\mu_{k_{l-1}}^{-1}}{\vert\mathcal{V}(k_{l-1})\vert} +
    \mu_j^{-1} +  \sum_{l=2}^{\ell'}\frac{\mu_{m_l}^{-1}-\mu_{m_{l-1}}^{-1}}{\vert\mathcal{V}(m_{l-1})\vert}.
\end{equation}
Introducing the nonnegative \emph{link weights} $w_{nm} = \frac{\mu_{n}^{-1}-\mu_m^{-1}}{\vert\mathcal{V}(m)\vert}$ if $m$ is a non-leaf nodes and $w_{nm}=\mu_n^{-1}$ if $m$ is a leaf node, and denoting by $\mathcal{P}(i,j) = \lbrace (k_0,k_1),(k_1,k_2),$ \dots$,(k_\ell,m_{\ell'}),\dots,(m_2,m_1),(m_1,m_0)\rbrace$ the \emph{shortest undirected path} between $i$ and $j$ in $\mathcal{T}$, expression \eqref{eqA11: node-based expression} simplifies to %[ref had to introduce weird break in the line with $\mathcal{P}$]
\begin{equation*}
    \omega_{ij} = \sum_{(m,n)\in\mathcal{P}(i,j)}w_{mn},
\end{equation*}
where the righthandside equals the weighted shortest path distance between $i$ and $j$ in $\mathcal{T}$. This proves Theorem \ref{thm4: Effective resistance}.$\hfill\square\medskip$

%%%%%%%%%%%%%
% REFERENCES
%%%%%%%%%%%%%
\bibliographystyle{siamplain}
\bibliography{Bibliography}
%\bibliography{references} AS IN EXAMPLE PACKAGE
\end{document}